\documentclass[twocolumn]{aastex631}

\usepackage{amsmath}
\usepackage{mathrsfs}
\usepackage{tensor}

\graphicspath{{./}{figures/}}

\definecolor{ao(english)}{rgb}{0.0, 0.5, 0.0}

\begin{document}

\title{Physics of Pair Producing Gaps in Black Hole Magnetospheres: Two Dimensional General Relativistic Particle-in-cell Simulations}

\correspondingauthor{Yajie Yuan}
\email{yajiey@wustl.edu}

\author[0000-0002-0108-4774]{Yajie Yuan}
\affil{Physics Department and McDonnell Center for the Space Sciences, Washington University in St. Louis; MO, 63130, USA}

\author[0000-0002-4738-1168]{Alexander Y. Chen}
\affil{Physics Department and McDonnell Center for the Space Sciences, Washington University in St. Louis; MO, 63130, USA}

\author[0009-0004-3333-7897]{Martin Luepker}
\affil{Physics Department and McDonnell Center for the Space Sciences, Washington University in St. Louis; MO, 63130, USA}

\begin{abstract}

Black holes can launch powerful jets through the Blandford-Znajek process. This relies on enough plasma in the jet funnel to conduct the necessary current. However, in some low luminosity active galactic nuclei, the plasma supply near the jet base may be an issue. It has been proposed that spark gaps---local regions with unscreened electric field---can form in the magnetosphere, accelerating particles to initiate pair cascades, thus filling the jet funnel with plasma. In this paper, we carry out 2D general relativistic particle-in-cell (GRPIC) simulations of the gap, including self-consistent treatment of inverse Compton scattering and pair production. We observe gap dynamics that is fully consistent with our earlier 1D GRPIC simulations. We find strong dependence of the gap power on the soft photon spectrum and energy density, as well as the strength of the horizon magnetic field. We derive physically motivated scaling relations, and applying to M87, we find that the gap may be energetically viable for the observed TeV flares. For Sgr A$^*$, the energy dissipated in the gap may also be sufficient to power the X-ray flares.

\end{abstract}


\section{Introduction} \label{sec:intro}
It is widely believed that black holes can launch powerful jets through the Blandford-Znajek process \citep{1977MNRAS.179..433B}. This is a promising mechanism to explain the relativistic jets powered by active galactic nuclei (AGN) and accreting X-ray binaries. When a rotating black hole is threaded by a coherent magnetic field, its spacetime drags the field lines into rotation, twisting them and launching an out-flowing Poynting flux. This process can only happen when there is enough plasma around the black hole so that the field is frozen into the plasma, or equivalently, there is a sufficient amount of charges to conduct the necessary current. However, charges in the jet funnel are continuously depleted: they either fall into the black hole when they are close to the event horizon, or get flung out due to centrifugal effect at large distances. 
Therefore, a mechanism to continuously replenish the plasma in the jet funnel is required for the Blandford-Znajek process to operate.

There are several ways this may happen. The first one is the mixing of materials from the disk into the jet funnel. This likely requires efficient plasma instabilities at the interface between the jet and the disk \citep[e.g.,][]{2021ApJ...914...55W}. It is not yet well understood whether the mass entrainment is fast enough to supply sufficient plasma all the way to the center of the jet. 
So far in 3D general relativistic magnetohydrodynamics (GRMHD) simulations, usually some artificial plasma source needs to be introduced in the jet to maintain the plasma supply (e.g. a density floor).
The second possibility is that the disk or its corona may emit a large number of MeV photons; they propagate into the jet funnel and collide with each other to create pairs---the so-called drizzle pair production \citep[e.g.,][]{2011ApJ...730..123L,2011ApJ...735....9M,2021ApJ...907...73W}. This may produce enough pairs to sustain the jets in high luminosity sources. However, in low luminosity sources, like M87 or Sgr A$^{*}$, drizzle pair production may not be enough. When both processes fail, a third mechanism can still fill the jet funnel with plasma: as the charge density runs low, the rotation induced electric field will have a component that is parallel to the magnetic field, which can accelerate charges to high energies
to collide with soft photons near the black hole to initiate pair production cascades \citep[e.g.,][]{1977MNRAS.179..433B}, which is able to replenish the plasma in the jet funnel. The regions with non-ideal electric field are called \emph{gaps}.

The magnetospheric gaps have been proposed to explain the fast $\gamma$-ray variability observed from low luminosity AGN, including M87 \citep[e.g.,][]{2011ApJ...730..123L, 2014Sci...346.1080A, 2017ApJ...841...61A, 2018ApJ...852..112K}. M87 exhibits powerful flares in the very high energy (VHE) $\gamma$-ray band, with isotropic equivalent luminosity reaching $L_{\rm VHE}\sim 10^{42}\,{\rm erg\,s^{-1}}$ \citep{2012ApJ...746..151A}. The total jet power is estimated to be $L_j\sim 10^{44}\,{\rm erg\,s^{-1}}$, so $L_{\rm VHE}\sim 0.01 L_j$. The variability time scale can be as short as one day; given the black hole mass $M=6.5\times10^9M_{\odot}$ \citep{2019ApJ...875L...1E}, the light crossing time of the gravitational radius is $r_g/c=GM/c^3\sim9$ hr, so the emitting region is likely not much larger than a few $r_g$ in size. In addition, some flares have been observed to be accompanied by increased radio emission from the nucleus \citep{2009Sci...325..444A}, suggesting that efficient particle acceleration may indeed happen near the black hole. The magnetospheric gap has been considered as one of the most promising mechanisms for producing the VHE $\gamma$-ray flares in M87 \citep{2011ApJ...730..123L,2018ApJ...852..112K}.

There has been a long history of analytical studies of the gap physics \citep{1992SvA....36..642B, 1998ApJ...497..563H,  2015ApJ...809...97B, 2016A&A...593A...8P, 2016ApJ...818...50H, 2016ApJ...833..142H, 2017ApJ...845...77H, 2018PhRvD..98f3016F, 2017PhRvD..96l3006L}. However, the analytical approach encounters its limits when treating the fully time-dependent nature of the gap and the plasma microphysics. Recently, general relativistic particle-in-cell (GRPIC) simulations have enabled first-principles studies of the magnetospheric gaps. \citet{2018A&A...616A.184L}, \citet{2018ApJ...863L..31C}, and \citet{2020ApJ...895..121C} first carried out 1D simulations of the pair discharge process including realistic radiative processes. In particular, \citet{2020ApJ...895..121C} observed macroscopic gaps opening and screening quasiperiodically near the so-called null surface in Boyer-Lindquist coordinates---the place where the zero angular momentum observer (ZAMO) measured charge density is zero. They also obtained physically motivated scaling relations and deduced that for M87 parameters, the dissipated power in the gap can reach 0.1\% of the jet power. \citet{2019PhRvL.122c5101P} was the first to carry out 2D GRPIC simulations of black hole magnetospheres. They studied jet launching when the black hole is immersed in an asymptotically uniform magnetic field, but with an ad hoc prescription for pair injection. \citet{2020PhRvL.124n5101C} used 2D GRPIC simulations with realistic treatment of inverse Compton (IC) scattering and pair production to study the pair discharge process around a spinning black hole threaded by a monopolar magnetic field. They reported a different picture compared to the 1D simulations of \citet{2020ApJ...895..121C}: highly time-dependent spark gaps open near the inner light surface (viewed in the Kerr-Schild coordinates) and inject pair plasma into the magnetosphere. Recently, the spark gaps in black hole magnetospheres have been further studied using 1D \citep{2020ApJ...902...80K,2024ApJ...964...78K} and 2D \citep{2023MNRAS.526.2709N} GRPIC simulations, confirming the time dependent nature of the gaps, and the necessity of pair injection between the two light surfaces.

Despite the significant progress, the existence of qualitative differences between 1D \citep{2020ApJ...895..121C,2020ApJ...902...80K} and 2D \citep{2020PhRvL.124n5101C} simulations calls for more detailed studies of the gap physics. In this work, we use our new GPU-based GRPIC code framework \emph{Aperture} \citep{2025arXiv250304558C} to carry out 2D global simulations of the pair producing gaps in a monopolar magnetosphere of a rapidly rotating black hole. We describe our simulation setup in \S\ref{sec:setup}. We then present the detailed results, including measured scaling relations in \S\ref{sec:results}. We discuss the physics behind the scaling relations and the implications for realistic astrophysical systems in \S\ref{sec:discussion}. Our conclusions are summarized in \S\ref{sec:conclusion}.

\section{Simulation setup} \label{sec:setup}
We 
start with a Kerr black hole in a vacuum monopolar magnetic field. In what follows, we use the Kerr-Schild spherical coordinates, and adopt the $3+1$ formalism following \citet{2004MNRAS.350..427K} (see Appendix \ref{sec:coord_initial} and \ref{sec:equations} for more details).
The vacuum field has nonzero $\mathbf{D}\cdot\mathbf{B}$, namely, there is a nonzero electric field parallel to the magnetic field, so any charge present near the black hole will be accelerated by this parallel electric field. We also assume the existence of a soft radiation field surrounding the black hole. The accelerated charges can go through inverse Compton scattering on the soft photons to produce gamma rays, and the gamma rays may collide with the soft photons to produce electron/positron pairs\footnote{The pair production due to collisions between IC generated gamma rays is negligible, because their number density is much lower than the soft photons in the situations we consider.}. This allows pair cascades to develop. For simplicity, we neglect synchrotron radiation and curvature radiation for the moment. We assume that the soft photon field is isotropic and uniform in the Kerr-Schild FIDO frame, with a gray body spectrum --- the same spectral shape as black body but the energy density is an independent variable beside the temperature. We write the energy density of the radiation field in terms of the photon energy $\epsilon$ as follows
\begin{equation}\label{eq:gray_body_spectrum}
    I(\epsilon)=\frac{n_0}{2(k T)^3 \zeta(3)}\frac{\epsilon ^3}{\exp\left(\epsilon/k T\right)-1},
\end{equation}
and the number density as a function of the photon energy is $n(\epsilon)=I(\epsilon)/\epsilon$. Here $n_0$ is the total number density of the soft photons, which we assume to be independent of $r$ and $\theta$; $T$ is the temperature, $k$ is the Boltzmann constant, and $\zeta$ is the Riemann zeta function. The average photon energy is $\langle\epsilon\rangle=k T \pi^4/(30\zeta(3))\approx2.7 k T$.

In our numerical simulations, we use a unit system such that lengths are measured in units of the gravitational radius $r_g$, times in units of $r_g/c$, and magnetic field in terms of the cyclotron frequency: $\tilde{B}=eBr_g/(m_ec^2)$, where the tilde denotes the dimensionless quantity, $e$ is the electron charge, $m_e$ is the electron mass, and $c$ is the speed of light. Similar to \citet{2020ApJ...895..121C}, we find that there are three dimensionless parameters in the problem. The first one is
\begin{equation}
    \tilde{B}_0=\frac{eB_0r_g}{m_ec^2},
\end{equation} 
where $B_0$ is the characteristic magnetic field strength at the event horizon. $\tilde{B}_0$ is the ratio between the cyclotron frequency and the inverse of the light crossing time of the gravitational radius $r_g$. Meanwhile, noticing that the characteristic Goldreich-Julian charge density can be written as 
\begin{equation}
    \rho_0=\frac{B_0}{4\pi r_g},
\end{equation}
the plasma skin depth corresponding to this is
\begin{equation}
    \lambda_p=\sqrt{\frac{m_ec^2r_g}{eB_0}},
\end{equation}
so we have $\tilde{B}_0=(r_g/\lambda_p)^2$. Therefore, $\tilde{B}_0$ sets the ratio between macroscopic and microscopic length scales.

The second dimensionless parameter is the temperature of the background photon field in units of electron rest mass, $\tilde{\epsilon}_0=kT/(m_ec^2)$, which is important in determining the IC scattering regimes and pair production threshold. The third parameter is
\begin{equation}
    \tau_0=r_g n_0 \sigma_T,
\end{equation}
which is the characteristic optical depth to
inverse Compton scattering in the Thomson regime. It determines the efficiency of IC scattering and pair production. As we will find out later \citep[and consistent with][]{2020ApJ...895..121C}, only the parameter combination $\tilde{B}_0\tilde{\epsilon}_0$ and $\tau_0$ matter for the gap dynamics. 

To determine a reasonable range for the dimensionless numerical parameters, we look at two supermassive black holes where the spark gap may operate. For M87, the horizon magnetic field is estimated to be $B_0\sim 10^2\,\mathrm{G}$ as required by the jet power \citep[e.g.,][]{2015ApJ...809...97B} \footnote{The Event Horizon Telescope observation of M87 polarized emission \citep{2021ApJ...910L..13E} suggests that the magnetic field is $B\sim1-30$ G at an emission radius of $r\sim 5r_g$. Assuming $B\propto1/r$, the magnetic field at the horizon is $B_0\sim 5-150$ G. Our estimation of $B_0\sim 10^2$ G is consistent with this.}
and since $r_g=9.6\times10^{14}$ cm for a supermassive black hole of mass $M=6.5\times10^9M_{\odot}$, we have $\tilde{B}_0\sim 6\times10^{13}$. The soft radiation near the black hole peaks in the mm range. If we assume a gray body spectrum, then $k T\sim 1$ meV, so $\tilde{\epsilon}_0\sim 2\times 10^{-9}$. The combination $\tilde{B}_0\tilde{\epsilon}_0$ is thus on the order of $10^5$. To get an estimation of $\tau_0$, we notice that the Event Horizon Telescope measures a total compact flux density of 0.64 Jy at 1.3 mm \citep{2019ApJ...875L...4E}; assuming most of the flux comes from a region of radius $r\sim 2r_g$ near the black hole \citep{2021MNRAS.507.4864Y}, we get an energy density of the soft photon field $u_s\sim0.04\,{\rm erg\, cm^{-3}}$ at $r\sim 2r_g$. This would give a $\tau_0\sim 6\times10^3$.  

For Sgr A$^{*}$, the black hole mass is $M=4.0_{-0.6}^{+1.1}\times 10^6M_{\odot}$ and the distance to us is $D=8.15\pm0.15$ kpc \citep{2022ApJ...930L..15E}. \citet{2022ApJ...930L..16E} estimated the magnetic field to be $B\sim 29$ G at $r\sim 5r_g$, so the horizon magnetic field is on the order of $B_0\sim 1.5\times10^2$ G. Given the gravitational radius $r_g=GM/c^2=6\times 10^{11}$ cm, we have $\tilde{B}_0\sim 5\times 10^{10}$. The soft photon field near the event horizon also peaks in the mm range, similar to M87, so $\tilde{\epsilon}_0\sim 2\times 10^{-9}$. As a result, $\tilde{B}_0\tilde{\epsilon}_0\sim 10^2$. \citet{2022ApJ...930L..13E} measured a time-averaged flux density of 2.4 Jy at 230 GHz. We can then estimate the energy density of the soft photon field at $r\sim 2r_g$ to be $u_s\sim0.08\,{\rm erg\, cm^{-3}}$, and the optical depth $\tau_0\sim 8$.  

We use our new GPU-based GRPIC code framework \emph{Aperture} \citep{2025arXiv250304558C} to simulate the pair discharge process in the aforementioned setup. We carry out 2D axisymmetric simulations, with a grid uniformly spaced in the $\theta$ direction from $\theta=0$ to $\theta=\pi$ and logarithmically spaced in $r$ from slightly below the event horizon to $r\sim 9r_g$. Our initial condition has a vacuum magnetic monopole field (see Appendix \ref{sec:coord_initial} for more details), and the space surrounding the black hole is filled with a plasma whose FIDO-measured density is proportional to $\sin\theta/(r\sqrt{\gamma})$ and the spatial velocity of individual particles is $u_j=0$.  We apply absorbing boundary conditions at the outer boundary to prevent waves and plasma from getting back into the simulation domain from larger radii. To properly resolve the plasma scales, namely the gyro radius and plasma skin depth, we have to scale down $\tilde{B}_0$, while maintaining  $\tilde{B}_0\tilde{\epsilon}_0\gg1$ and $\tilde{\epsilon}_0\ll1$. Our fiducial runs have $\tilde{B}_0=10^4$ and $\tilde{\epsilon}_0=0.02$, so $\tilde{B}_0\tilde{\epsilon}_0=200$, and a typical resolution $6144\times6144$. Our highest resolution run is $12288\times12288$ for the case of $\tilde{B}_0=2\times10^4$. We employ fully self-consistent IC and pair production treatment in \emph{Aperture} \citep{2020ApJ...895..121C}. We probe a series of $\tau_0$ ranging from 1 to 200. Our simulation parameters resemble the condition near Sgr A*, while some scaling is required to extrapolate our results to M87 parameters. In all the simulations, the dimensionless spin parameter of the black hole is $a=0.999$.

\section{Results} \label{sec:results}

\subsection{Overall qualitative dynamics}\label{subsec:overall}

\begin{figure*}
    \centering
    \includegraphics[width=\textwidth]{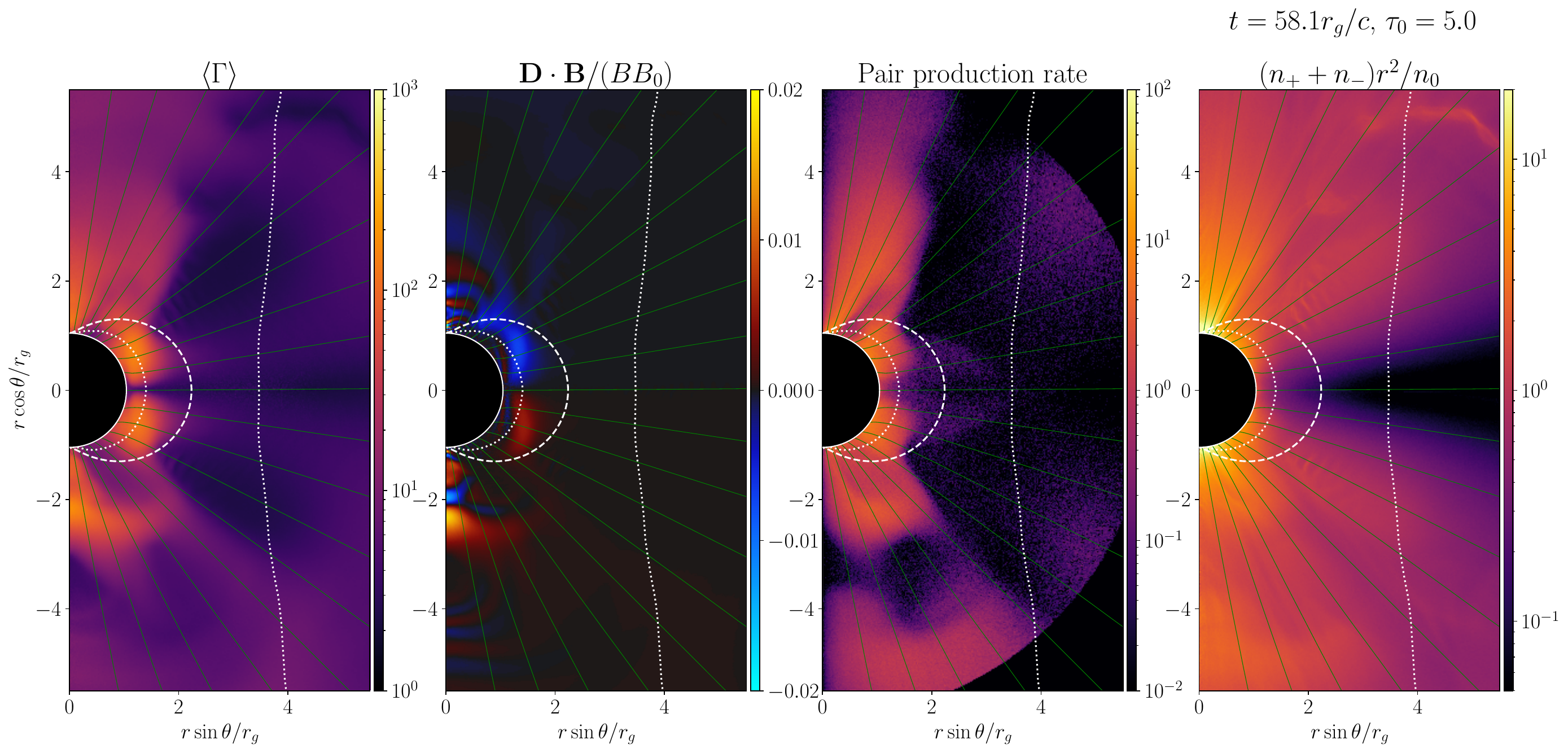}\\
    \includegraphics[width=\textwidth]{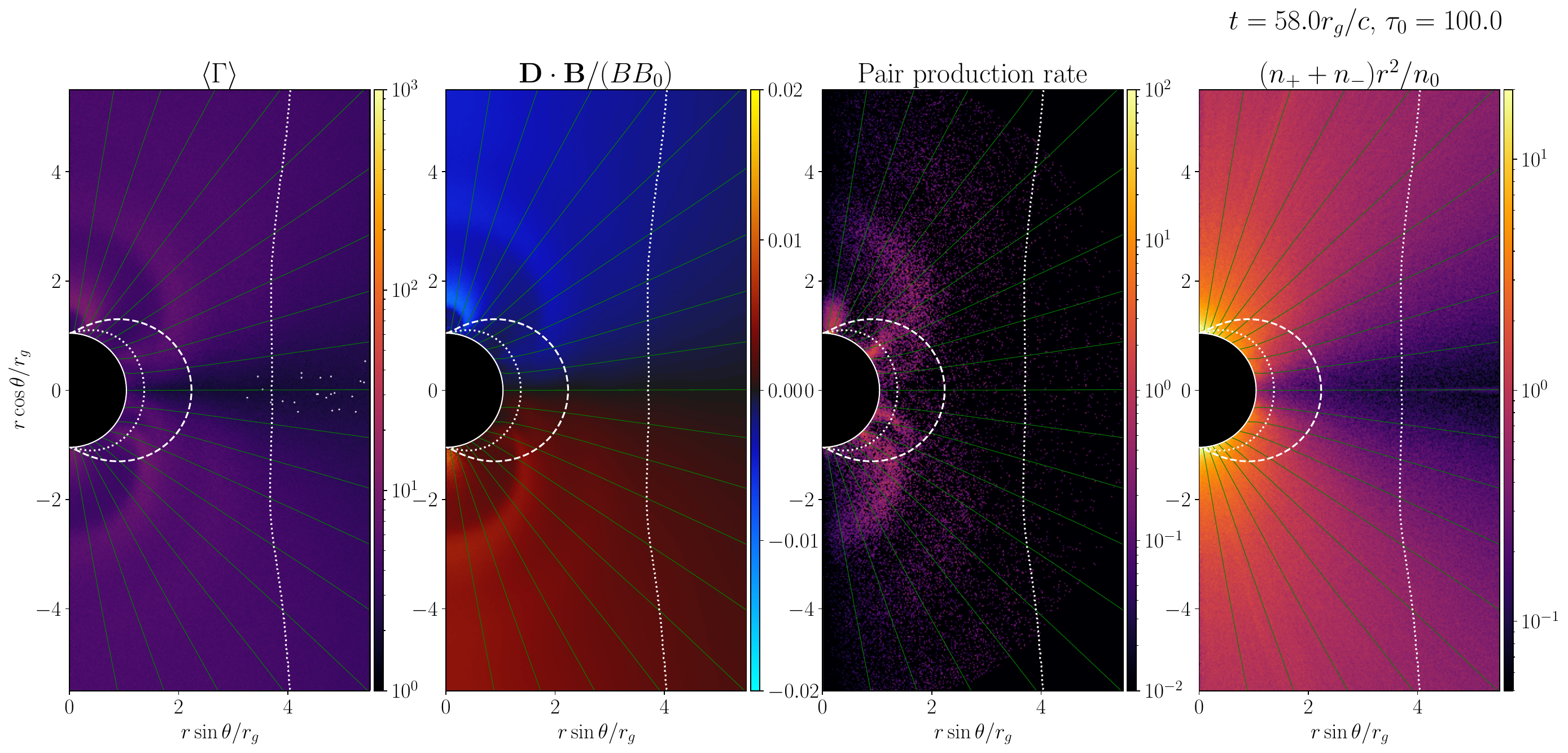}
    \caption{Snapshots from two simulations with different optical depths, $\tau_0=5$ (top) and $\tau_0=100$ (bottom). In both simulations, $\tilde{B}_0=10^4$ and $\tilde{\epsilon}_0=0.02$ are kept the same. From left to right: average Lorentz factor of particles in the FIDO frame $\langle \Gamma\rangle$, the electric field component parallel to the magnetic field $\mathbf{D}\cdot\mathbf{B}/(B B_0)$, pair production rate (we plot the value of $r\sqrt{-g}\, (dS^0/dt)$ in arbitrary units, where $S^0$ is the zeroth component of the number-flux-4-vector), and the plasma density. Green lines are magnetic field lines. The white dashed line is the ergosphere, and the white dotted lines are the light surfaces.}
    \label{fig:snapshot}
\end{figure*}

We first focus on a series of simulations where we set $\tilde{B}_0=10^4$, $\tilde{\epsilon}_0=0.02$. We observe qualitatively different behaviors of the magnetosphere when we change the optical depth $\tau_0$. 

When $\tau_0\lesssim 3$, the electric field parallel to the magnetic field grows quickly as the initial plasma is depleted from the magnetosphere but not enough pairs are produced. The growing electric field accelerates particles into the Klein-Nishina regime, further reducing the IC and pair production cross-section. This leads to a runaway growth of the electric field. The end result is that the magnetosphere becomes vacuum-like. 

When $3<\tau_0\lesssim 20$, we see bursts of pair production that screen the electric field parallel to the magnetic field, and the magnetosphere becomes approximately force-free. As the plasma gets gradually depleted, a few coherent regions with nonzero parallel electric field develop near the event horizon. These are the gaps where particles are accelerated and pair cascades are initiated. Figure \ref{fig:snapshot} top row shows snapshots when the gaps are in action from a simulation with $\tau_0=5$. In this case, we see gaps that are coherent across a wide angular range, reaching half of a hemisphere or even the whole hemisphere in the most coherent case. The gap electric field is antiparallel to the background magnetic field in the northern hemisphere, and parallel to the background magnetic field in the southern hemisphere. This is consistent with the fact that the monopole field is pointing outward, so the current in the force-free magnetosphere flows out of the black hole from the southern hemisphere and flows into the black hole in the northern hemisphere. With the Kerr-Schild time slicing, we see the gaps start near the event horizon, gradually moving outward; the screening happens behind this front, in the form of a large amplitude oscillation of the parallel electric field. After the screening, charges will gradually leave the magnetosphere, and new gaps open up again. This process happens quasiperiodically. The pair production rate has a significant angular dependence. There is not much pair production near the equator, but the rate increases at higher latitudes. This is because the current density goes to zero at the equator, and increases toward the poles. Note that the outgoing Poynting flux follows the opposite trend: it is maximum at the equator, and goes to zero at the poles. This suggests that the pair production process is in fact driven by the current.

When $\tau_0$ further increases beyond $\sim 20$, the parallel electric field cannot be fully screened, even though bursts of pair production still happen. The magnetosphere reaches a steady state with a significant amount of residual parallel electric field. The bottom row of Figure \ref{fig:snapshot} shows snapshots of the steady state in a simulation with $\tau_0=100$. We see frequent, quasiperiodic bursts of pair production happening, screening/reducing the parallel electric field near the event horizon, as evident in the second and third panel. However, the pairs are not enough to screen the parallel electric field at increasingly large distances. The average Lorentz factors of particles remain low, as shown in the first panel. This is likely due to the strong radiative drag in the high $\tau_0$ regime. In this particular case shown in Figure \ref{fig:snapshot}, the residual parallel electric field is much less than the vacuum one, but we find it to increase with increasing $\tau_0$. A more quantitative study will be given in \S\ref{subsec:scaling}.

\begin{figure}
    \centering
    \includegraphics[width=\columnwidth]{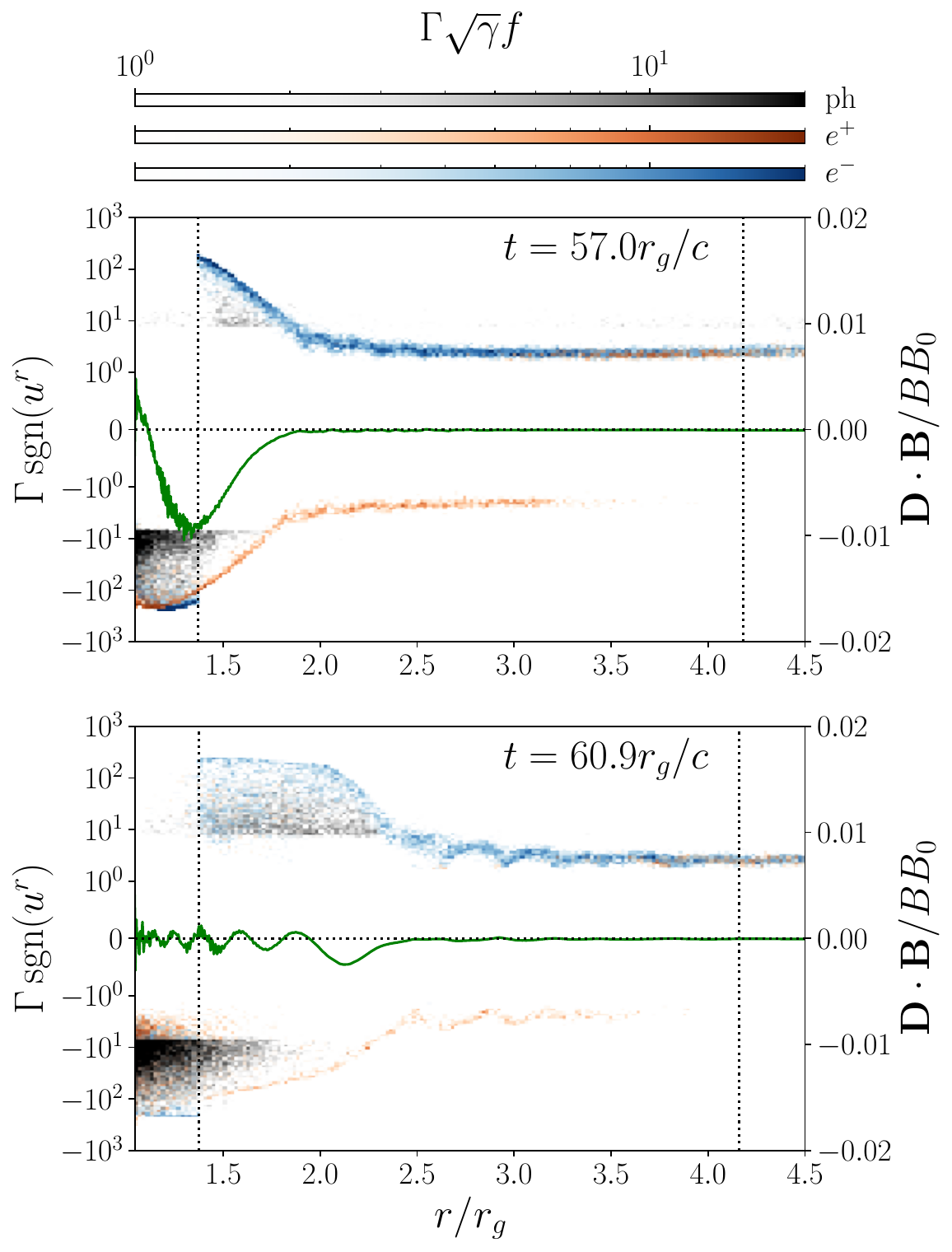}
    \caption{The phase space distribution of electrons (blue), positrons (orange) and photons (gray) along a radial line at $\theta=60^{\circ}$, at two different times during a cycle of gap opening and screening in our fiducial run (the same simulation as shown in the top row of Figure \ref{fig:snapshot}). The energy distribution is in terms of the particle Lorentz factor $\Gamma$ in the FIDO frame (for photons, we use $\epsilon/(m_ec^2)$, where $\epsilon$ is the photon energy in the FIDO frame), and we separate outgoing particles and ingoing particles using the sign of the radial component of their 4-velocity $u^r$. Also plotted is the normalized parallel electric field $\mathbf{D}\cdot\mathbf{B}/B B_0$ (the green line). The vertical dotted lines mark the locations of the light surfaces. The left boundary of the horizontal axis corresponds to the event horizon. Note that we only track photons with $\epsilon/(m_ec^2)>5$.}
    \label{fig:1D_phase}
\end{figure}

\subsection{A detailed look at the physics of the gap}

\begin{figure}
    \centering
    \begin{minipage}{\columnwidth}
    \begin{center}
        (a) Kerr-Schild coordinates
    \end{center}
    \vspace{-0.3cm}
    \includegraphics[width=\columnwidth]{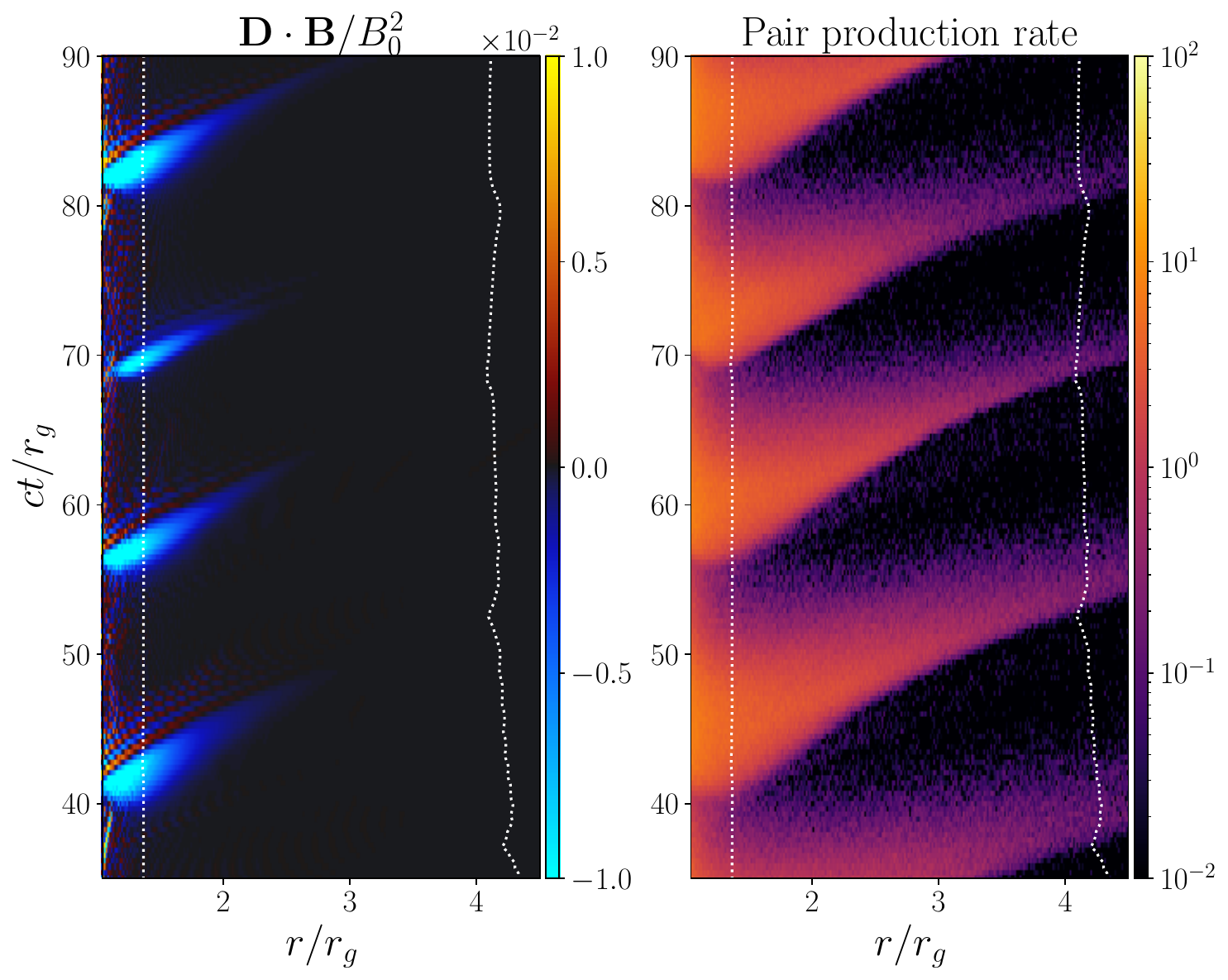}
    \end{minipage}
    \begin{minipage}{\columnwidth}
    \vspace{0.3cm}
    \begin{center}
        (b) Boyer-Lindquist coordinates
    \end{center}
    \vspace{-0.3cm}
    \includegraphics[width=\columnwidth]{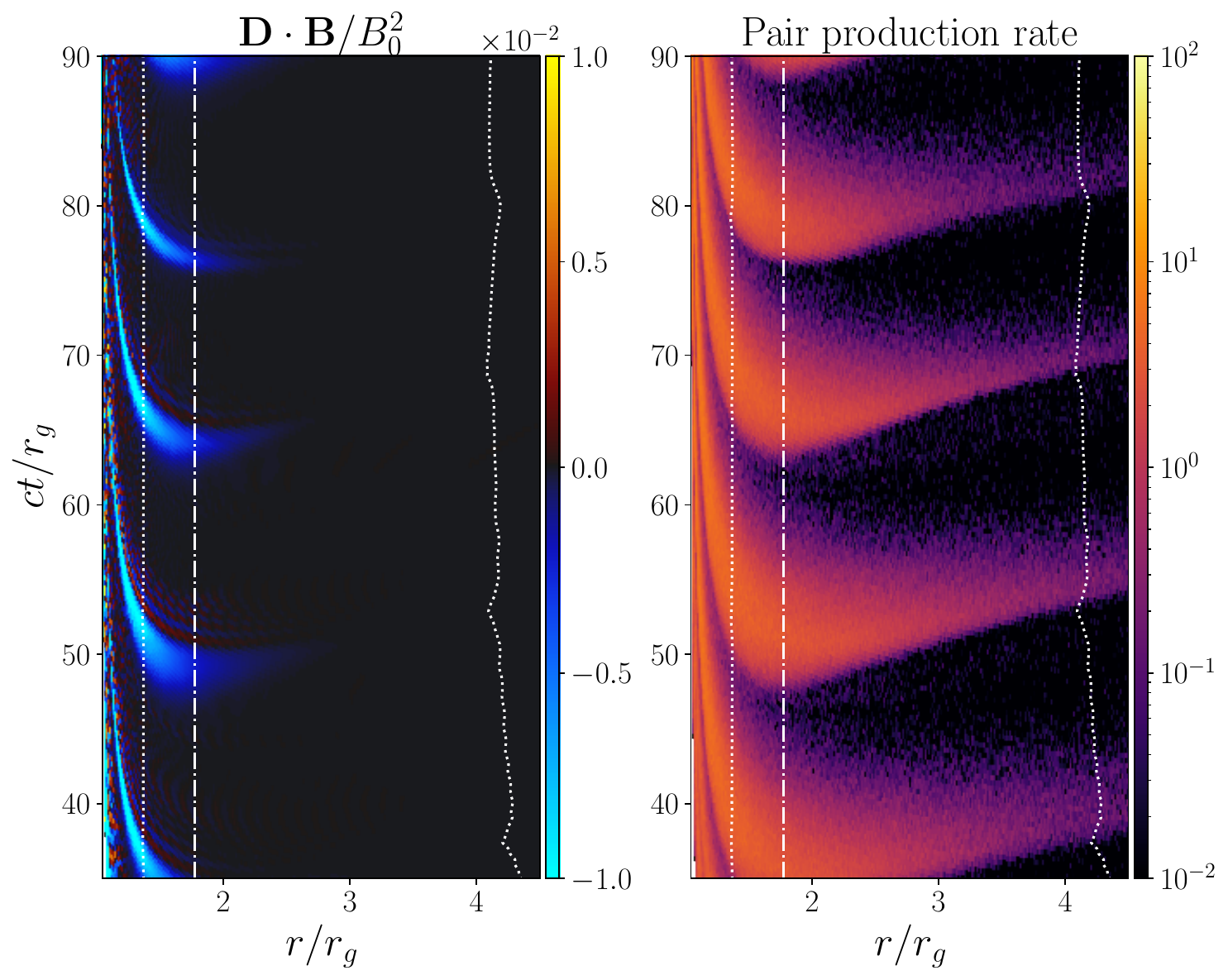}
    \end{minipage}
    \caption{Spacetime diagram of $\mathbf{D}\cdot\mathbf{B}/B_0^2$ (left) and the pair production rate $r\sqrt{-g}\, (dS^0/dt)$ (right), along a radial line at $\theta=60^{\circ}$, for our fiducial run. The top row is in Kerr-Schild coordinates, and the bottom row in Boyer-Lindquist coordinates. Both quantities are invariant between the two coordinate systems. The white dotted lines mark the location of the light surfaces. In the bottom row, the null surface is shown as the white dash-dotted line. The left boundary of the horizontal axis corresponds to the event horizon.}
    \label{fig:spacetime}
\end{figure}

\begin{figure*}
    \centering
    \includegraphics[width=0.7\textwidth]{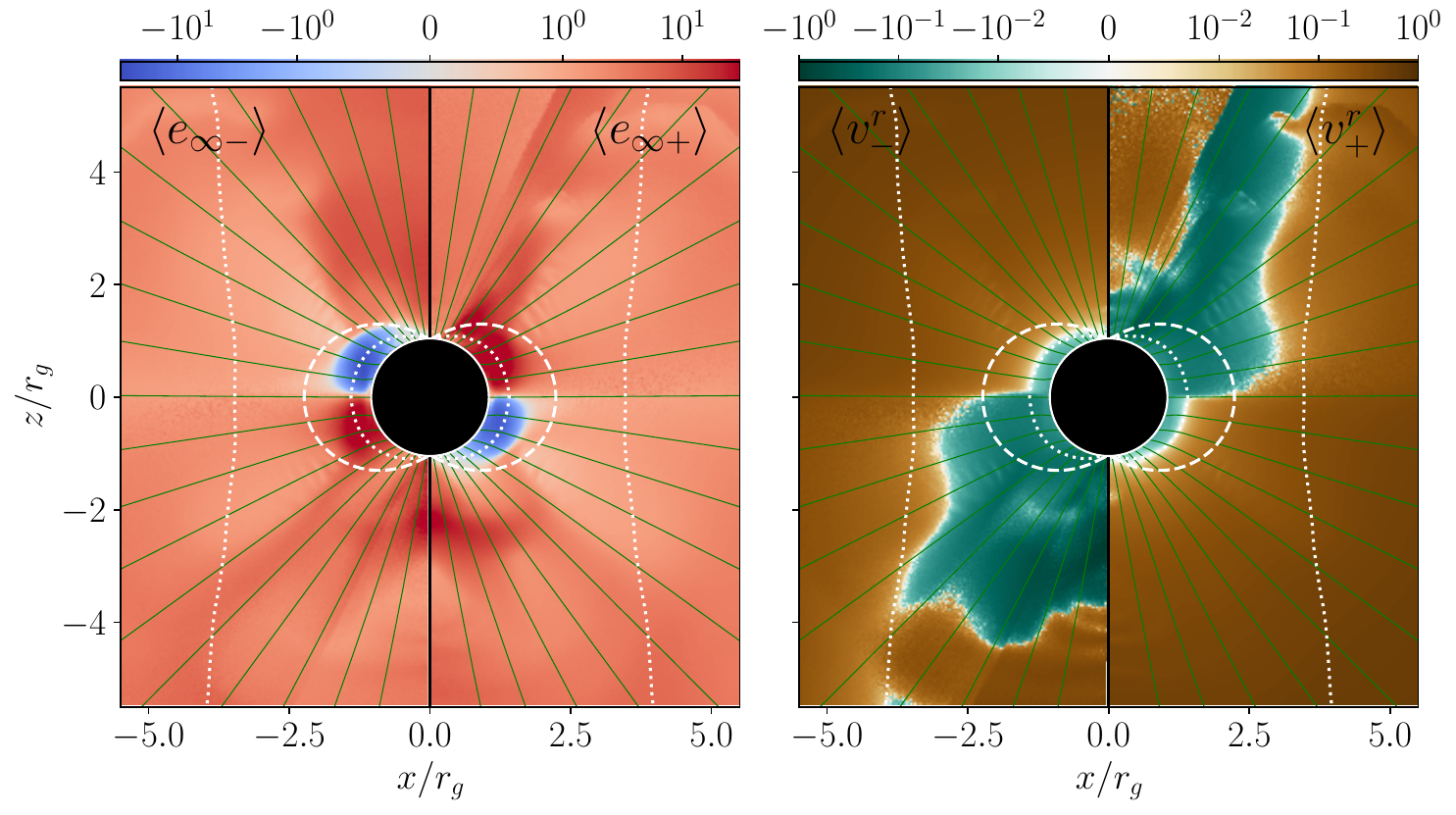}
    \caption{Average particle energy at infinity and radial velocity in our fiducial run, for electrons ($-$) and positrons ($+$), at the same time as in Figure \ref{fig:snapshot} top row. The white dashed lines and white dotted lines indicate the ergosphere and the light surfaces, respectively.}
    \label{fig:einf_vr}
\end{figure*}

\begin{figure}
    \begin{minipage}{\columnwidth}
    \begin{center}
        (a) Power as a function of radius at $t=58.1r_g/c$
    \end{center}
    \vspace{-0.3cm}
    \includegraphics[width=\columnwidth]{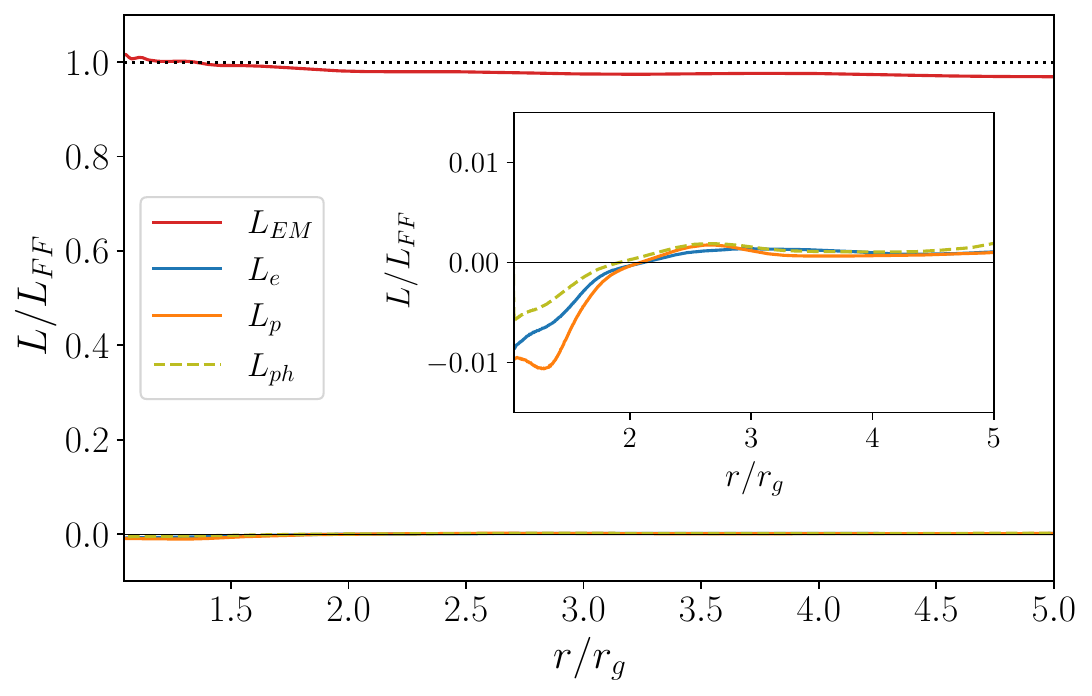}
    \end{minipage}
    \begin{center}
        (b) Power as a function of time at two radii
    \end{center}
    \vspace{-0.3cm}
    \begin{minipage}{\columnwidth}
    \includegraphics[width=\columnwidth]{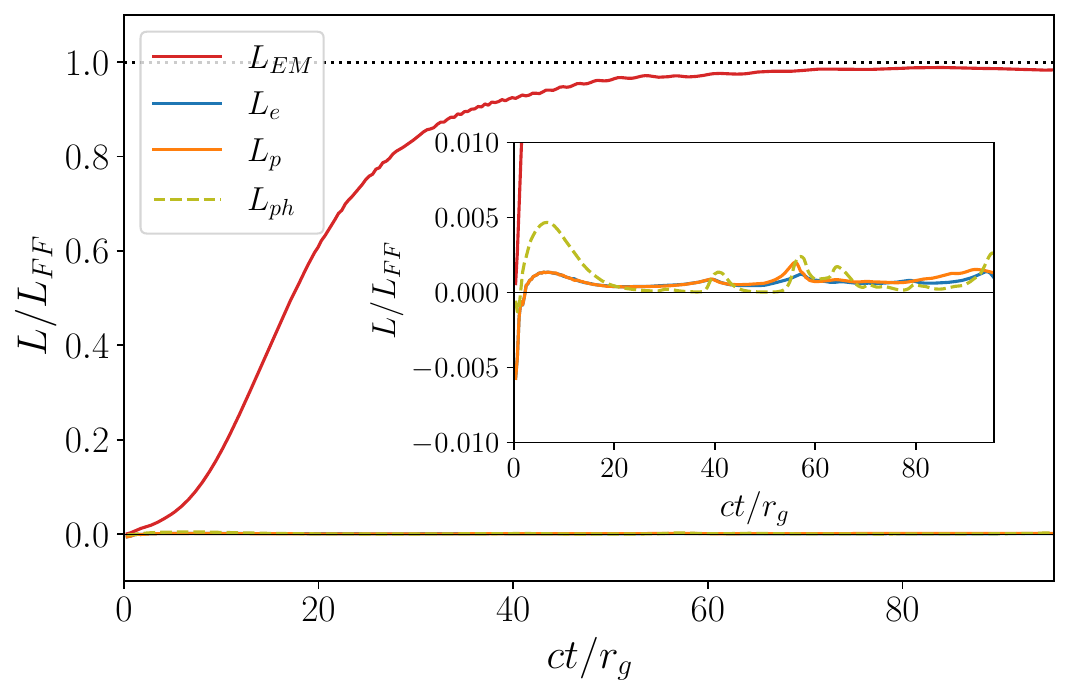}
    \includegraphics[width=\columnwidth]{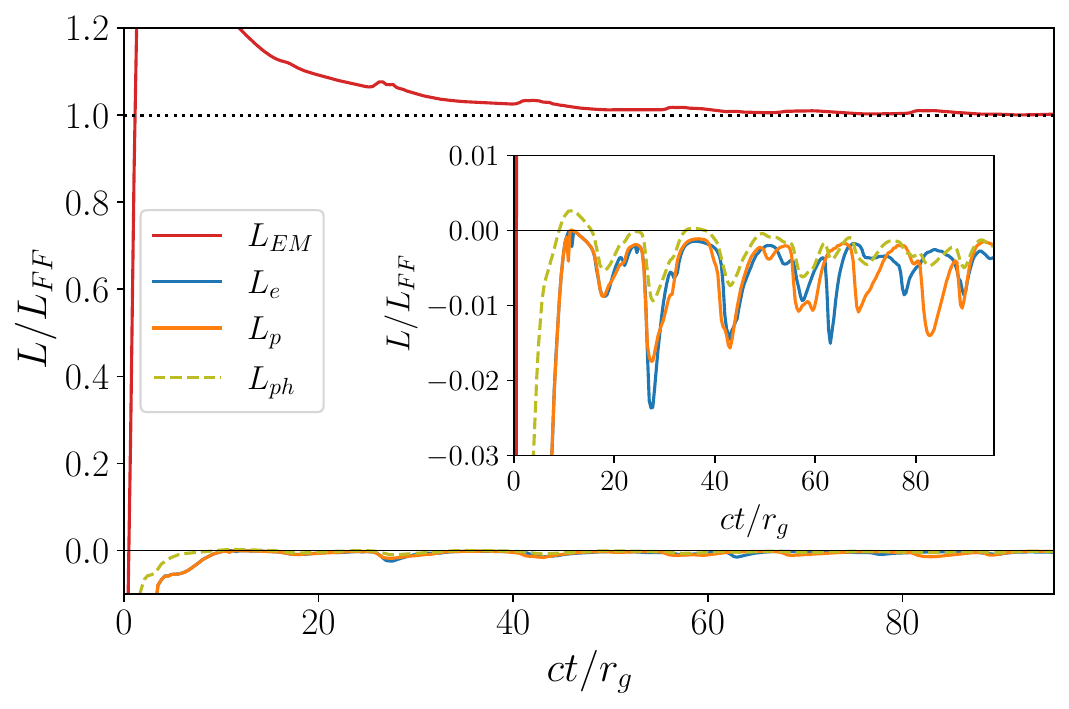}
    \end{minipage}
    \caption{The total power going through spherical surfaces in our fiducial run, decomposed into different components: $L_{EM}$---electromagnetic power, $L_e$---power carried by electrons, $L_p$---power carried by positrons, $L_{ph}$---power carried by tracked high energy photons with $\epsilon/(m_ec^2)>5$. (a) The power as a function of radius, at the same time as in Figure \ref{fig:snapshot} top row. (b) The power going through the surface at $r=5r_g$ (upper panel) and through the event horizon (lower panel), respectively, as a function of time. The power is positive if it is going outward. It is normalized by the force-free value $L_{FF}$. The insets show zoom-in views near $L=0$.}
    \label{fig:tau5_Lr_Lt}
\end{figure}

In what follows, we carry out detailed analysis of the physics of the gap, for our fiducial run with $\tilde{B}_0=10^4$, $\tilde{\epsilon}_0=0.02$, and $\tau_0=5$, where we see quasiperiodic gap opening and screening.

We first focus on the spatial and temporal structure of the gaps. Figure \ref{fig:1D_phase} shows two phase space snapshots of the one-dimensional slice along the radial direction at a polar angle $\theta=60^{\circ}$, during 
one of the gap screening cycles. In the top panel, the gap has opened up, and its electric field has reached the maximum. We see that the parallel electric field accelerates electrons and positrons in opposite directions, pushing them to high Lorentz factors. Although the gap itself can straddle the inner light surface, the particles cannot move outward radially when they are inside the inner light surface. The jump of the typical value of $\Gamma\, {\rm sgn}(u^r)$ for electrons at the inner light surface indicates that the electron velocity has a significant non-radial component (mostly $\phi$ component). Similarly, particles do not move inward radially when they are outside the outer light surface. In the bottom panel, we see that the right front of the gap has moved outward, and the gap electric field has been almost screened, only some large amplitude oscillations remaining. The fraction of high energy electrons and positrons is decreasing as they lose energy through IC scattering, and new pairs at lower energies are produced.

In Figure~\ref{fig:spacetime} top row, we show the spacetime diagram of $\mathbf{D}\cdot\mathbf{B}/B_0^2$ and the pair production rate along the same radial line at $\theta=60^{\circ}$ in Kerr-Schild coordinates, the same coordinates in which we carry out our simulations. The quasiperiodicity of the gap shows up very clearly in the spacetime diagram. From the left panel, we see that the gap electric field first develops near the event horizon, then moves outward, with an apparent radial velocity $\sim 0.2c$. Similarly, for the pair production rate, it first sets in near the event horizon, then expands to larger radii.

As has been pointed out by \citet{2025arXiv250304558C}, the appearance of the spacetime diagram depends on the coordinate system used. In Figure \ref{fig:spacetime} bottom row, we show the spacetime diagram along the same radial line, in Boyer-Lindquist coordinates. Although the $r$ and $\theta$ coordinates remain the same between Kerr-Schild and Boyer-Lindquist coordinates, $t$ and $\phi$ are different (see equations \ref{eq:BL_KS_global_t}--\ref{eq:BL_KS_global_phi}). In particular, given the same time in Kerr-Schild coordinates $t_{\rm KS}$, the corresponding time in Boyer-Lindquist coordinates $t_{\rm BL}$ stretches out more towards the future as we get closer to the event horizon. As a result, in Boyer-Lindquist coordinates, the gap electric field appears to first develop somewhere in between the inner light surface and the outer light surface, then the two ends of the gap expand in both directions. We find that the location where the gap first develops is consistent with the null surface in Boyer-Lindquist coordinates---the surface where the corresponding FIDO (also the zero angular momentum observer, or ZAMO frame) measured charge density is zero in the force-free solution. The null surface is overlaid on the spacetime diagram in the bottom panel of Figure \ref{fig:spacetime} (the null surface is obtained from our semi-analytical force-free solution, see Appendix \ref{sec:force-free}). The screening of the gap also starts near the null surface, then expands both ways. Note that the null surface is a frame/coordinates-dependent feature; in Kerr-Schild coordinates, the FIDO measured charge density in the force-free solution never changes sign, so there is no null surface. What we observe here in
2D is consistent with earlier 1D studies of \cite{2020ApJ...895..121C} and \cite{2020ApJ...902...80K}, who performed their simulations in Boyer-Lindquist coordinates.

Next, we turn our focus to the physics of energy extraction from the black hole. To see whether the Penrose process \citep{Penrose:1969pc} plays an important role, we first look for particles with negative energy at infinity $e_{\infty}=-u_0$, where $u_0$ is the time component of the covariant 4-velocity of the particle. Figure \ref{fig:einf_vr} left panel shows $\langle e_{\infty}\rangle$ for electrons and positrons, respectively, at a particular time when the gap is in action. We do see negative energy particles. In particular, electrons with $e_{\infty}<0$ exist in the northern hemisphere near the event horizon and within the ergosphere, while positrons with $e_{\infty}<0$ exist in a more or less symmetric region in the southern hemisphere. The regions with $e_{\infty}<0$ particles exist throughout the simulation, but they have the largest spatial extent and the highest $|\langle e_{\infty}\rangle|$ values when the gap is active near the event horizon. It is the electric field that tries to push the electrons or positrons away from the black hole that puts the particles on the negative energy orbit. In the other hemisphere, the electric field pushes the particles toward the black hole, and the particle energy at infinity remains positive. 

The right panel of Figure \ref{fig:einf_vr} shows that the average radial velocities of particles are indeed toward the black hole within the inner light surface. In the northern hemisphere, electrons with $e_{\infty}<0$ fall into the black hole, so energy is extracted from the black hole through the Penrose process; however, the positrons that fall into the black hole has positive $e_{\infty}$, depositing energy into the black hole. The opposite happens in the southern hemisphere. To determine the net effect, we calculate the power going through spherical surfaces carried by different components, including the electromagnetic field, electrons, positrons, and high energy photons (see Appendix \ref{sec:averaging} for the details of the calculation). The results are shown in Figure \ref{fig:tau5_Lr_Lt}.

We can see that in this fiducial run, during the steady state, the electromagnetic power is very close to the force-free value at all radii. The power carried by electrons, positrons and photons is very small, at most a few percent. At the event horizon, the net energy flux of the particles is actually going into the black hole. In this case, the energy extraction from the black hole is primarily through the Blandford-Znajek process rather than the Penrose process.

We find that even in the cases with large $\tau_0$, where the parallel electric field is not fully screened, the energy flux in the particles is still negligibly small compared to the Poynting flux. Near the event horizon, the net particle energy flux is still into the black hole. In these cases, the overall Poynting flux can significantly drop below the force-free value.

\subsection{Scaling relations}\label{subsec:scaling}

\begin{figure}
    \centering
    \includegraphics[width=\columnwidth]{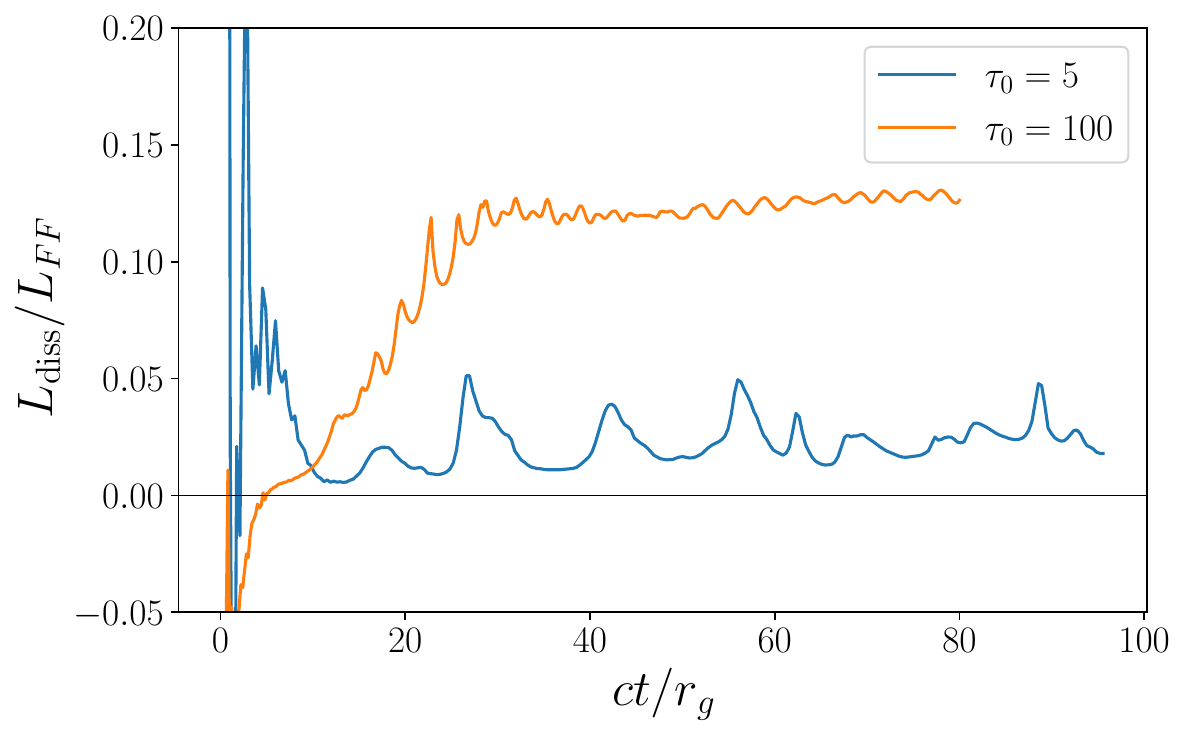}
    \caption{The power dissipated in the gap as a function of time, for two runs with the same $\tilde{B}_0=10^4$ and $\tilde{\epsilon}_0=0.02$, but different $\tau_0$.}
    \label{fig:Ldiss_time}
\end{figure}

\begin{figure}
    \centering
    \includegraphics[width=\columnwidth]{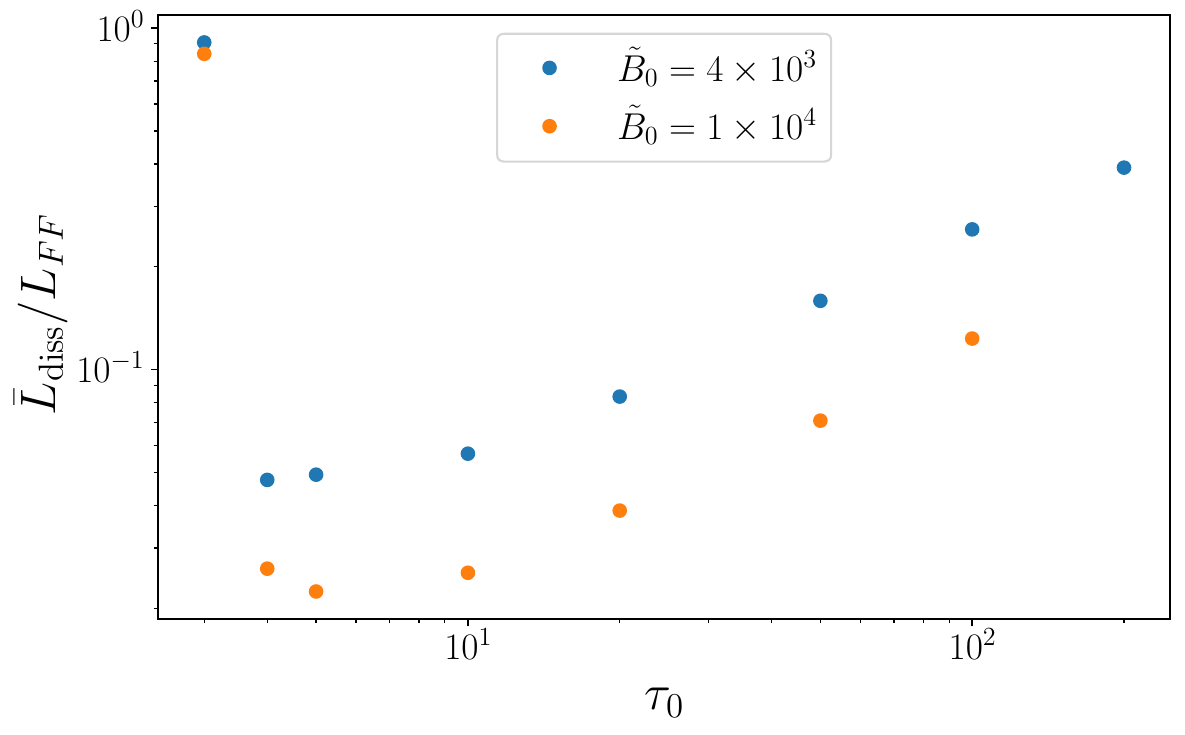}
    \includegraphics[width=\columnwidth]{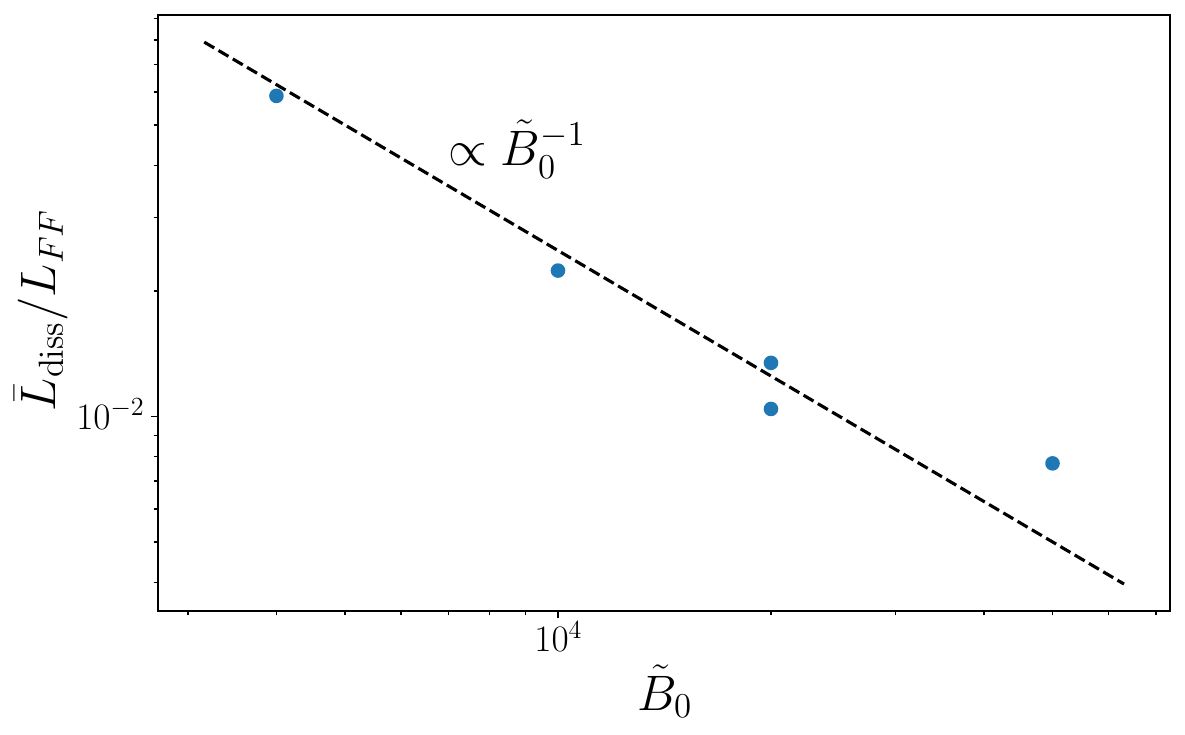}
    \caption{Scaling of the time averaged dissipation power during the steady state, $\bar{L}_{\rm diss}$, with $\tau_0$ (top panel) and $\tilde{B}_0$ (bottom panel). In all the simulations, we have $\tilde{\epsilon}_0=0.02$. In the top panel, two data series with different $\tilde{B}_0$ are shown. In the bottom panel, all simulations have $\tau_0=5$.}
    \label{fig:scaling}
\end{figure}

We would like to quantify the amount of power dissipated in the gap and how it scales with the dimensionless parameters, including $\tau_0$ and $\tilde{B}_0$. We calculate the dissipation power as:
\begin{align}\label{eq:L_diss}
    L_{\rm diss}&=\int_{\rm all\ space} \mathbf{E}\cdot\mathbf{j}\sqrt{\gamma}\,dr\,d\theta\,d\phi. 
\end{align}
Figure \ref{fig:Ldiss_time} shows $L_{\rm diss}$ as a function of time for two runs with the same $\tilde{B}_0$ and $\tilde{\epsilon}_0$, but different $\tau_0$. For the $\tau_0=5$ case, during the steady state, we see peaks in $L_{\rm diss}$ that correspond to active gaps. When the gaps are screened, $L_{\rm diss}$ returns to a relatively low value. For the $\tau_0=100$ case, the overall dissipation level is higher, due to the significant residual parallel electric field that cannot be screened. The bursts of pair production also imprint peaks/oscillations on the $L_{\rm diss}$ curve.

We calculate the time averaged dissipation power during the steady state, $\bar{L}_{\rm diss}$, to compare across different simulations. Figure \ref{fig:scaling} shows the results. In the top panel, we compare simulations where $\tilde{B}_0$ and $\tilde{\epsilon}_0$ are held the same, while $\tau_0$ varies. $\bar{L}_{\rm diss}$ is minimum at $\tau_0\sim4-5$, and this remains true for the different magnetic field values $\tilde{B}_0$ we have simulated. For $\tau_0\gtrsim5$, $\bar{L}_{\rm diss}$ increases with $\tau_0$. Generally, we see gap cycles getting shorter and the residual parallel electric field becoming more significant as $\tau_0$ increases. This leads to enhanced dissipation in the gap. For $\tau_0\lesssim 3$, the parallel electric field cannot be screened and the magnetosphere becomes vacuum like. The high value of $\bar{L}_{\rm diss}$ is due to the initial particles in the simulation domain that are accelerated by the electric field to high energies.

In the bottom panel of Figure \ref{fig:scaling}, we compare $\bar{L}_{\rm diss}$ for simulations with different $\tilde{B}_0$, while other parameters are kept the same. The dissipation power goes down as $\tilde{B}_0$ increases, roughly following $\bar{L}_{\rm diss}\propto \tilde{B}_0^{-1}$. In our simulations with higher $\tilde{B}_0$, we generally observe smaller gaps, and they are less coherent across the angular range. 

\section{Discussion}\label{sec:discussion}

\subsection{Physics behind the scaling relations}\label{subsec:scaling_analytical}
To understand the dependence of the gap power on the optical depth $\tau_0$, we develop a semi-analytical model for the electric field in the gap.

\begin{figure}
    \centering
    \includegraphics[width=\columnwidth]{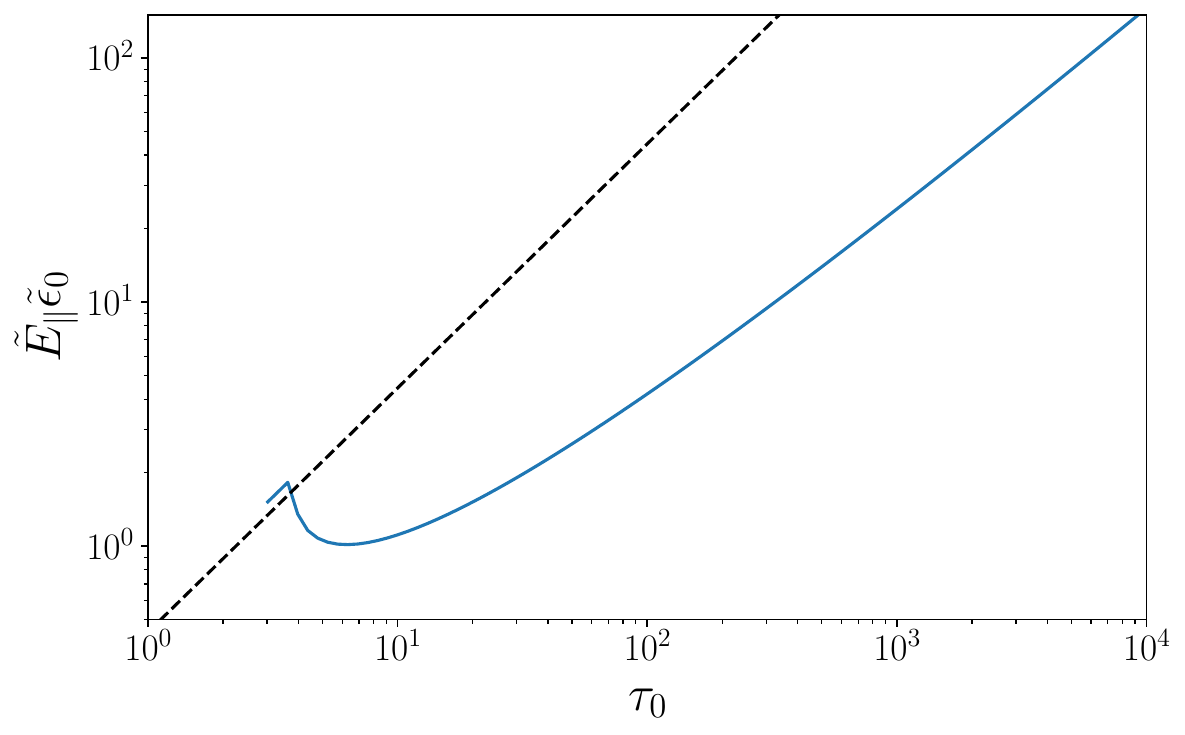}
    \includegraphics[width=\columnwidth]{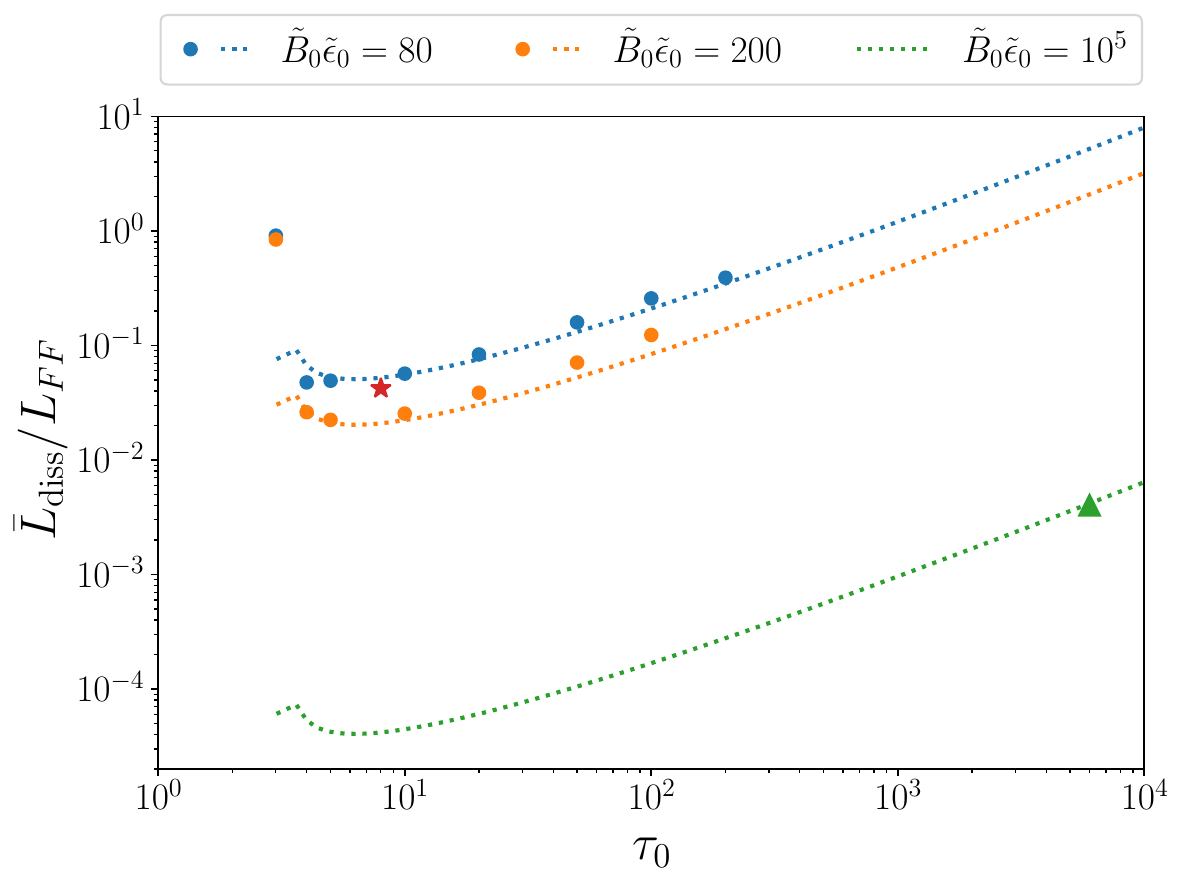}
    \caption{Top panel: the blue solid line shows the minimum electric field needed for the particles accelerated in the gap to produce pairs, namely, Equation (\ref{eq:f_def}). The black dashed line shows the electric field that will accelerate particles into the Klein-Nishina regime, as in Equation (\ref{eq:E_KN}). Bottom panel: the derived scaling relation for the dissipation power, Equation (\ref{eq:L_diss_scaling}), plotted against the simulation data in the top panel of Figure \ref{fig:scaling}. The blue and orange dots are the data points, at $\tilde{B}_0\tilde{\epsilon}_0=80,\ 200$, respectively, and the dotted lines are the corresponding analytical scaling relations. We also show the extrapolation to M87 parameters, namely, $\tilde{B}_0\tilde{\epsilon}_0=10^5$, $\tau_0=6\times10^3$, which gives $L_{\rm diss}/L_{FF}=4\times 10^{-3}$, shown as the green triangle. For Sgr A$^*$ parameters, $\tilde{B}_0\tilde{\epsilon}_0=10^2$, $\tau_0=8$, we have $L_{\rm diss}/L_{FF}=0.04$, which is indicated by the red star.}
    \label{fig:extrapolate}
\end{figure}

We will first consider IC scattering in the Thomson regime---in particular, we refer to the regime where the interaction between the accelerated particles and the \emph{peak} of the soft photon spectrum is Thomson scattering. Suppose that when the gap electric field has developed to initiate the pair cascade, the primary particles are accelerated to a Lorentz factor $\gamma_p$ such that the acceleration is balanced by IC cooling:
\begin{equation}
    e E_{\parallel}=\frac{4}{3}\gamma_p^2\sigma_T n_0 \langle\epsilon\rangle=\frac{4}{3}\gamma_p^2 \frac{\tau_0}{r_g} \langle\epsilon\rangle.
\end{equation}
This gives
\begin{equation}\label{eq:gamma_p}
    \gamma_p=\sqrt{\frac{3e E_{\parallel} r_g}{4\tau_0\langle\epsilon\rangle}}.
\end{equation}
A primary particle escapes the region after a time scale $\sim r_g/c$. During this time, it will scatter a number of $N(\epsilon)$ soft photons at energy $\epsilon$:
\begin{align}
    N(\epsilon)&=r_g \sigma_T n(\epsilon)\epsilon 
    =\frac{\tau_0}{2(k T)^3 \zeta(3)}\frac{\epsilon^3}{\exp\left(\epsilon/k T\right)-1},
\end{align}
where we have assumed that the spectrum of the background soft photons follows the gray body distribution described by Equation~\eqref{eq:gray_body_spectrum}.
In the Thomson regime, these soft photons will be upscattered to an energy
\begin{equation}
    E_{\rm ph}\simeq 2\gamma_p^2\epsilon.
\end{equation}
These high energy photons will produce pairs with the soft photons, and the largest cross section $\sigma_{\gamma\gamma}\sim\sigma_T/5$ is near the threshold, namely, when a photon with energy $E_{\rm ph}$ interacts with soft photons of energy $\epsilon'=2(m_ec^2)^2/E_{\rm ph}$. As a result, the $N(\epsilon)$ high energy photons will produce a number of $\kappa(\epsilon)$ pairs:
\begin{align}
    \kappa(\epsilon)&=N(\epsilon)r_g\frac{\sigma_T}{5}[n(\epsilon')\epsilon']_{\epsilon'=2(m_ec^2)^2/E_{\rm ph}} \nonumber\\
    &=\frac{1}{5}\left(\frac{\tau_0 (m_ec^2)^3}{2\gamma_p^3(kT)^3\zeta(3)}\right)^2\frac{1}{\exp\left(\epsilon/kT\right)-1}\nonumber\\
    &\times\frac{1}{\exp\left[(m_ec^2)^2/\left(\gamma_p^2\epsilon kT\right)\right]-1}
\end{align}
This takes the maximum value at around $\epsilon=(m_ec^2)^2/(\gamma_p^2\epsilon)$, namely, $\epsilon=m_ec^2/\gamma_p$---this is borderline between Thomson regime and Klein-Nishina regime. The maximum value of $\kappa(\epsilon)$ is 
\begin{align}
    \kappa_m&=\frac{1}{5}\left(\frac{\tau_0 (m_ec^2)^3}{2\gamma_p^3(kT)^3\zeta(3)}\right)^2\frac{1}{\left[\exp\left(\frac{m_ec^2}{\gamma_p kT}\right)-1\right]^2}\nonumber\\
    &=\frac{16\tau_0^5\langle\epsilon\rangle^3 (m_ec^2)^6}{135(eE_{\parallel}r_g)^3(kT)^6\zeta(3)^2}\nonumber\\
    &\times\left[\exp\left(\frac{m_ec^2}{kT}\sqrt{\frac{4\tau_0\langle\epsilon\rangle}{3eE_{\parallel}r_g}}\right)-1\right]^{-2},
\end{align}
where we have plugged in Equation (\ref{eq:gamma_p}) to get the second equality.
Converting to dimensionless variables, and noticing that $\langle\tilde{\epsilon}\rangle\approx 3 \tilde{\epsilon}_0$, we get
\begin{equation}\label{eq:kappa_m_dimensionless}
    \kappa_m=\frac{16\tau_0^5}{5(\tilde{E}_{\parallel}\tilde{\epsilon}_0)^3\zeta(3)^2}\left[\exp\left(\sqrt{\frac{4\tau_0}{\tilde{E}_{\parallel}\tilde{\epsilon}_0}}\right)-1\right]^{-2}.
\end{equation}
We use $\kappa_m$ as a reasonable estimate for the number of pairs produced by one primary particle. In order for the gap to be screened, we need $\kappa_m\ge1$. This gives a lower bound on the parallel electric field needed. The solution of $\tilde{E}_{\parallel}$ for different optical depth $\tau_0$ when $\kappa_m=1$ in Equation (\ref{eq:kappa_m_dimensionless}) can be found numerically. We define
\begin{equation}\label{eq:f_def}
    (\tilde{E}_{\parallel}\tilde{\epsilon}_0)_{\kappa_m=1}\equiv f(\tau_0).
\end{equation}
The function $f(\tau_0)$ is shown as the blue line in the top panel of Figure \ref{fig:extrapolate}. Below $\tau_0\sim3-4$, a proper solution does not exist, as particles will always enter the Klein-Nishina regime. This is consistent with what we found in our PIC simulations. $f(\tau_0)$ is minimum around $\tau_0\sim 5-8$, and then grows monotonically as $\tau_0$ further increases. In the meantime, the typical Lorentz factor of the primary particles, $\gamma_p\tilde{\epsilon}_0=\sqrt{\tilde{E}_{\parallel}\tilde{\epsilon}_0/(4\tau_0)}$, gradually decreases as $\tau_0$ increases to large values, suggesting that the primary particles can no longer reach the optimal energy for pair production due to the strong Compton drag; they are mainly interacting with soft photons on the 
Wien
tail of the gray body distribution to produce pairs.

For each $\tau_0$ there is also an upper bound on the electric field. If the electric field is too strong, the IC scattering of the primary particles will be in the Klein-Nishina regime. This happens if $\gamma_p\langle\epsilon\rangle\gtrsim m_ec^2$, or the electric field exceeds the following value \citep{2019ApJ...883...66P}
\begin{equation}
    E_{\parallel}^{({\rm KN})}=\frac{4\tau_0(m_e c^2)^2}{3e r_g \langle\epsilon\rangle}.
\end{equation}
The dimensionless version is
\begin{equation}\label{eq:E_KN}
    \tilde{E}_{\parallel}^{({\rm KN})}=\frac{4\tau_0}{9\tilde{\epsilon}_0}.
\end{equation}
This is shown as the black dashed line in the top panel of Figure \ref{fig:extrapolate}. We can see that the lower bound of the electric field $f(\tau_0)$ is below this Klein-Nishina line except for very small $\tau_0$. This validates our assumption behind the derivation of $f(\tau_0)$.
Note that both in Equation (\ref{eq:kappa_m_dimensionless}) and (\ref{eq:E_KN}), $\tilde{E}_{\parallel}$ only appears in the combination $\tilde{E}_{\parallel}\tilde{\epsilon}_0$. This again confirms that $\tilde{\epsilon}_0$ only sets the energy scale in the problem; the physics does not depend on its absolute value. Therefore, only the combination $\tilde{E}_{\parallel}\tilde{\epsilon}_0$ or $\tilde{B}_{0}\tilde{\epsilon}_0$ matters.

The total dissipation power $L_{\rm diss}$ is the spatial integral of $\mathbf{E}\cdot\mathbf{j}$, as calculated in Equation (\ref{eq:L_diss}). For simplicity, we assume that the electric field is the lower bound calculated in Equation (\ref{eq:f_def}), which is independent of $\tilde{B}_0$, and we ignore the dependence of the gap size on $\tilde{B}_0$. Since the current is close to the force-free value, which is proportional to $\tilde{B}_0$ (applicable when the gap electric field is sufficiently small, i.e. $E_{\parallel}\ll B_0$), and the total force-free power $L_{FF}\propto \tilde{B}_0^2$ (see Appendix \ref{sec:force-free}), we see that $L_{\rm diss}/L_{FF}\propto \tilde{B}_0^{-1}$. We find that the following function
\begin{equation}\label{eq:L_diss_scaling}
    \frac{L_{\rm diss}}{L_{FF}}=\frac{C_1 f(\tau_0)}{\tilde{B}_0\tilde{\epsilon}_0}
\end{equation}
with a constant $C_1=4$ agrees very well with our GRPIC simulation data, as shown in the lower panel of Figure~\ref{fig:extrapolate}.


We would like to apply the scaling relations to realistic astrophysical systems. For M87, taking the estimations from \S\ref{sec:setup}, if we assume that the soft photon field near the event horizon of M87 has a gray body spectrum peaking at $\tilde{\epsilon}_0\sim 2\times 10^{-9}$, with $\tau_0\sim 6\times10^3$, and $\tilde{B}_0\tilde{\epsilon}_0\sim 10^5$, then using Equation (\ref{eq:L_diss_scaling}), we find that $L_{\rm diss}/L_{FF}\sim 4\times 10^{-3}$. For Sgr A$^*$, we have $\tau_0\sim 8$, and $\tilde{B}_0\tilde{\epsilon}_0\sim 10^2$, so $L_{\rm diss}/L_{FF}\sim 0.04$. Both cases are shown in the lower panel of Figure~\ref{fig:extrapolate}.

\subsection{The significance of the soft photon spectrum}

In this work, we focused on a soft photon field with a gray body spectrum, and found that the dissipation power in the gap grows as the optical depth $\tau_0$ increases. On the other hand, \cite{2018ApJ...863L..31C} and \cite{2020ApJ...895..121C} used a soft photon field with a hard power law spectrum, and they reported decreasing dissipation as optical depth increases. This seems to suggest that the gap behavior depends critically on the properties of the soft photon field.

To better understand these results, we carry out a similar analysis as in \S\ref{subsec:scaling_analytical}, for a general power law distribution of soft photons. We write the photon energy spectrum as 
\begin{equation}
    I(\epsilon)=n_0 s \left(\frac{\epsilon}{\epsilon_{\min}}\right)^{-s},\quad \epsilon\geq \epsilon_{\min},
\end{equation}
and the number density as
\begin{equation}
    n(\epsilon)=n_0 s \frac{\epsilon^{-s-1}}{\epsilon_{\min}^{-s}},\quad \epsilon\geq \epsilon_{\min}.
\end{equation}
This is normalized such that $\int n(\epsilon)\,d\epsilon=n_0$. The average soft photon energy is then $\langle\epsilon\rangle=\epsilon_{\min}s/(s-1)$. We require $s>1$ in order for the photons near $\epsilon_{\min}$ to dominate the energy content of the soft photon field.

We still assume that the primary particles reach a state where acceleration balances IC cooling, and that IC scattering is in Thomson regime for the mean soft photon energy $\langle \epsilon\rangle$, so the Lorentz factor of the primary particles $\gamma_p$ is still given by Equation~\eqref{eq:gamma_p}. During a time scale $\sim r_g/c$, a primary particle will scatter the following number of soft photons at energy $\geq \epsilon$:
\begin{equation}
    N(\epsilon)=r_g \sigma_T n(\epsilon)\epsilon =\tau_0 s \left(\frac{\epsilon}{\epsilon_{\min}}\right)^{-s}.
\end{equation}
The up scattered photons have a typical energy of $E_{\rm ph}\simeq2\gamma_p^2\epsilon$, and they produce the following number of pairs
\begin{align}
    \kappa(\epsilon)&=N(\epsilon)r_g\frac{\sigma_T}{5}[n(\epsilon')\epsilon']_{\epsilon'=2/E_{\rm ph}} =\frac{\tau_0^2}{5}s^2\left(\frac{\gamma_p\epsilon_{\min}}{m_ec^2}\right)^{2s}\nonumber\\
    &=\frac{\tau_0^{2-s}}{5}s^{2-s}\left(\frac{3}{4}\tilde{\epsilon}_{\min}\tilde{E}_{\parallel}(s-1)\right)^s.
\end{align}
Keep in mind that $\kappa(\epsilon)$ is the number of pairs produced from one primary particle by IC scattering on soft photons with energy $\geq \epsilon$.  It can be seen that $\kappa(\epsilon)$ is actually independent of $\epsilon$, and it increases monotonically with $\tilde{E}_{\parallel}$. In order for the gap to be screened, we need $\kappa(\epsilon)$ to reach a certain value $\kappa_m\geq 1$. This gives a lower bound on the electric field
\begin{equation}
    \tilde{E}_{\parallel}\tilde{\epsilon}_{\min}=\frac{4 (5\kappa_m)^{1/s}s^{\frac{s-2}{s}}}{3(s-1)}\tau_0^{\frac{s-2}{s}}.
\end{equation}
Clearly, $\tilde{E}_{\parallel}\tilde{\epsilon}_{\min}$ decreases with $\tau_0$ when $1<s<2$, and increases with $\tau_0$ when $s>2$. \cite{2018ApJ...863L..31C} and \cite{2020ApJ...895..121C} used a spectral index $s=1.2$, consistent with the former case. On the other hand, 
the Wien tail of a gray body spectrum resembles the latter case with large $s$,
so we see the opposite trend instead in \S\ref{subsec:scaling}.

In reality, the soft photon spectrum near the event horizon of the AGN is highly uncertain. In \S\ref{subsec:scaling_analytical}, for M87, assuming a gray body spectrum, and an estimated energy density according to EHT observations, we got an average dissipation rate in the gap $L_{\rm diss}/L_{FF}\sim 4\times 10^{-3}$. Since the jet power in M87 is on the order of $10^{44}\,{\rm erg\,s^{-1}}$, if most of the dissipation goes into $\gamma$-rays, the corresponding $\gamma$-ray luminosity would be $4\times10^{41}\,{\rm erg\,s^{-1}}$. This is slightly higher than the estimation by \cite{2018ApJ...863L..31C} and \cite{2020ApJ...895..121C}, who assumed a hard power law for the soft photons, but is still slightly lower than the brightest VHE $\gamma$-ray flares from M87, which can reach an isotropic equivalent luminosity $L_{\rm VHE}\sim 10^{42}\,{\rm erg\,s^{-1}}$. However, the time variability of the soft photon field, the gap dynamics itself, and the Doppler beaming due to the jet plasma flow may lead to enhanced $\gamma$-ray emission. Given the sensitive dependence on the soft photon spectrum, we need to use more realistic models for the radiation field near the black hole, e.g. those from radiation GRMHD simulations \citep{2021MNRAS.507.4864Y}, to predict whether a flare is more likely to happen when the optical depth $\tau_0$ increases or decreases. On the other hand, if the $\gamma$-ray flares are indeed produced by the discharging gaps, we can in turn utilize the sensitive dependence on the soft photon field to constrain the property of the low energy radiation near the black hole using the $\gamma$-ray flares.

As to Sgr A$^*$, also assuming a gray body spectrum for the soft photon field, we estimated the average dissipation rate $L_{\rm diss}/L_{FF}\sim 0.04$. Sgr A$^*$ does not show any obvious jets; however, if the horizon-threading magnetic field does launch a jet through the Blandford-Znajek mechanism, the jet power would be \citep{2015ApJ...809...97B}
\begin{equation}
    L=\frac{k}{4\pi c}\Omega_H^2\Phi^2,
\end{equation}
where $k\approx0.045$ for parabolic jet geometry, $\Phi$ is the poloidal magnetic flux threading the black hole, and $\Omega_H$ is the angular velocity of the horizon. Using a horizon field of $B_0\sim 1.5\times10^2$ G as estimated in \S\ref{sec:setup}, we can obtain the jet power $L\sim k a^2 c \pi B_0^2r_g^2\sim3\times10^{37}a^2\,{\rm erg\,s}^{-1}$. Therefore, the average power dissipated in the gap would be $L_{\rm diss}\sim10^{36}\,{\rm erg\,s}^{-1}$, if the black hole spin is close to maximum. This is in line with the energetics of X-ray flares, whose typical luminosity in the 2-10 keV range is $L_X\sim(1-3)\times 10^{35}\,{\rm erg\,s}^{-1}$ \citep[e.g.,][]{2013ApJ...769..155D}. 
The soft photon field near the horizon of Sgr A$^*$ can be highly variable \citep{2022ApJ...930L..15E}. Our estimation of $\tau_0\sim8$ using the average 230 GHz flux is already very close to the threshold $\tau_{m}\sim3$, below which gap will no longer be screened. It is possible that the variability will result in $\tau_0$ evolving across $\tau_{m}$ back and forth, which may give highly variable dissipation in the gap, even intermittent jets. 
Another caveat is that we do not yet have a measurement of the black hole spin, for both Sgr A$^{*}$ and M87. In this work, we only simulated a black hole with close to maximal spin $a=0.999$. We leave it to future work to study the dependence of $L_{\rm diss}/L_{FF}$ on the black hole spin.

In this work, we neglected other radiative processes, e.g., synchrotron radiation and curvature radiation. These can be important energy loss channels for accelerated particles \citep{2020ApJ...902...80K,2022ApJ...924...28K,2024ApJ...964...78K}, and can also reshape the multi-wavelength spectrum. We will consider their effects in future works. In addition, we used a simplified field geometry, i.e. a monopole. Realistic fields may well have a current layer near the equator as the magnetic field changes sign going from one hemisphere to the other. It will be interesting to see which process dominates in the energy dissipation, the pair discharge gaps or the magnetic reconnection at the current sheet. We leave this to future studies.

\section{Conclusions} \label{sec:conclusion}
We have carried out 2D GRPIC simulations of the pair discharge process in a monopolar magnetosphere around a rapidly spinning Kerr black hole. We included realistic IC scattering and pair production on a gray body soft photon distribution. Our main conclusions are the following:
\begin{enumerate}
    \itemsep0em 
    \item The steady state of the magnetosphere depends strongly on the optical depth $\tau_0$ for IC scattering and pair production. For a gray body soft photon field, there is an optimal range, $\tau_0\sim 4-20$, where the electric field parallel to the magnetic field can be fully screened and the magnetosphere is nearly force-free. For $\tau_0\lesssim 3$, particles are accelerated into Klein-Nishina regime and the magnetosphere becomes vacuum like. For $\tau_0\gtrsim 20$, although pair production happens in bursts, there is a significant residual parallel electric field.
    \item For the cases where the parallel electric field can be screened, we do see macroscopic gaps open quasiperiodically. The appearance of the gap and how it is screened depend on the coordinate system. In Kerr-Schild coordinates, gaps
    start near the event horizon, then move outward as a result of the screening process.
    In Boyer-Lindquist coordinates, gaps first develop near the null surface.
    The screening also happens there, leading to two shrinking gaps moving in opposite directions away from the null surface.
    Our results are consistent with earlier 1D simulations by~\cite{2020ApJ...895..121C} and the 2D simulations with simplified pair production microphysics by~\citet{2025arXiv250304558C}.
    \item In all cases, the energy extraction from the black hole is primarily in the form of Poynting flux. Although we see negative energy particles moving into the black hole, the overall contribution of the particles carries energy into the black hole, and the fraction is small compared to the Poynting flux.
    \item We have measured and derived the scaling of the power dissipated in the gap with the relevant parameters, $\tau_0$ and $\tilde{B}_0\tilde{\epsilon}_0$. We find that the dissipation power scales inversely with $\tilde{B}_0\tilde{\epsilon}_0$, and the scaling with $\tau_0$ depends on the soft photon spectrum. In particular, for a gray body or a steep power law, the dissipation power increases with increasing $\tau_0$, but the trend is opposite for a hard power law.
    \item For M87, assuming that the soft photon field has a gray body spectrum peaking in the mm range with EHT measured flux, and horizon magnetic field $B_0\sim 10^2$ G, we estimate an average dissipation power in the gap to be $\sim4\times 10^{-3}$ times the jet power. This is lower than the brightest very high energy $\gamma$-ray flares, but the variability of the soft photon field may lead to higher dissipation. Further studies using more realistic soft photon field are needed.
    \item For Sgr A$^{*}$, we estimated that the average dissipation power in the gap may reach $10^{36}\,{\rm erg\,s}^{-1}$. This may be sufficient to power the X-ray flares from Sgr A$^{*}$. The highly variable mm emission near Sgr A$^{*}$ may lead to strong variation in the gap power and jet dynamics.
\end{enumerate}

\begin{acknowledgments}
We thank Sam Gralla, Amir Levinson, John Mehlhaff, Kyle Parfrey, and Sasha Philippov for helpful discussion.
AC and YY acknowledge support from NSF grants DMS-2235457 and AST-2308111. AC also acknowledges support from NASA grant 80NSSC24K1095. This work was also facilitated by the Multimessenger Plasma Physics Center (MPPC), NSF grant PHY-2206608, and by a grant from the Simons Foundation (MP-SCMPS-00001470) to YY. This research used resources of the Oak
Ridge Leadership Computing Facility at the Oak Ridge National Laboratory,
which is supported by the Office of Science of the U.S. Department of Energy
under Contract No. DE-AC05-00OR22725.
\end{acknowledgments}

\software{APERTURE \citep{2025arXiv250304558C}
          }

\appendix

\section{Coordinate system and initial condition}\label{sec:coord_initial}
We use the 3+1 formalism \citep{2004MNRAS.350..427K} in our numerical simulations. The metric can be written as
\begin{equation}
    ds^2=\left(\beta ^2-\alpha ^2\right)dt^2 +2 \beta _i dx^i dt+ \gamma _{ij}dx^i dx^j,
\end{equation}
where $\alpha$ is the lapse function, $\pmb{\beta}$ the shift vector, and $\gamma_{ij}$ the induced metric on the 3D spatial hypersurface of constant coordinate time $t$. The 4-velocity of the local fiducial observer (FIDO) is 
\begin{align}
    n_{\mu}&=(-\alpha,\mathbf{0}),\nonumber\\
    n^{\mu}&=\frac{1}{\alpha}(1,-\pmb{\beta}).
\end{align}
In what follows, we use the geometrized units with $G=c=1$ and the black hole mass $M=1$. In spherical Kerr-Schild coordinates $(t,r,\theta,\phi)$, the metric is
\begin{align}
ds^2=g_{\mu\nu}dx^{\mu}dx^{\nu}&=-\left(1-\frac{2 r}{\Sigma }\right)dt^2+\frac{4 r}{\Sigma} dt\, dr+\left(1+\frac{2 r}{\Sigma }\right)dr^2+\Sigma\, d\theta^2+\frac{A \sin ^2\theta}{\Sigma } d\phi^2\nonumber\\
&-\frac{4 a r \sin ^2\theta }{\Sigma }\, dt\, d\phi - 2 a \left(1+\frac{2 r}{\Sigma }\right) \sin ^2\theta\, dr\, d\phi,
\end{align}
where $\Sigma=r^2+a^2\cos^2\theta$, $A=(r^2+a^2)^2-\Delta a^2\sin^2\theta$, $\Delta=r^2-2 r+a^2$, and $a$ is the dimensionless spin parameter of the black hole. So we have
\begin{align}
    \alpha&=\frac{1}{\sqrt{1+2 r/\Sigma}},\\
    \beta^i&=(\frac{2 r}{\Sigma+2 r},0,0),\\
    \gamma_{ij}&=g_{ij}.
\end{align}
The spherical Kerr-Schild coordinates can be obtained from Boyer-Lindquist coordinates through the following transformation:
\begin{align}
    dt_{KS}&=dt_{BL}+\frac{2r}{\Delta}dr_{BL},\label{eq:BL_KS_diff_t}\\
    dr_{KS}&=dr_{BL},\label{eq:BL_KS_diff_r}\\
    d\theta_{KS}&=d\theta_{BL},\label{eq:BL_KS_diff_theta}\\
    d\phi_{KS}&=d\phi_{BL}+\frac{a}{\Delta}dr_{BL}.\label{eq:BL_KS_diff_phi}
\end{align}

Upon integration, we obtain the global coordinate transformation \citep[e.g.,][]{2010MNRAS.406.2047G}
\begin{align}
    t_{\rm KS}&=t_{\rm BL}+\frac{1}{\sqrt{1-a^2}}\left[\left(1+\sqrt{1-a^2}\right)\ln \left|\frac{r_{\rm BL}}{1+\sqrt{1-a^2}}-1\right|-\left(1-\sqrt{1-a^2}\right)\ln \left|\frac{r_{\rm BL}}{1-\sqrt{1-a^2}}-1\right|\right],\label{eq:BL_KS_global_t}\\
    r_{\rm KS}&=r_{\rm BL},\label{eq:BL_KS_global_r}\\
    \theta_{\rm KS}&=\theta_{\rm BL},\label{eq:BL_KS_global_theta}\\
    \phi_{\rm KS}&=\phi_{\rm BL}+\frac{a}{2\sqrt{1-a^2}}\ln\left|\frac{r_{\rm BL}-(1+\sqrt{1-a^2})}{r_{\rm BL}-(1-\sqrt{1-a^2})}\right|.\label{eq:BL_KS_global_phi}
\end{align}

The vacuum magnetic monopole solution is described by the following 4-vector potential in the Boyer-Lindquist coordinates \citep[e.g.,][]{crinquand_particle_2021}
\begin{equation}
    \mathscr{A}_{\mu}^{BL}=B_0\left(\frac{a \cos\theta}{\Sigma},0,0,-\cos\theta\,\frac{r^2+a^2}{\Sigma}\right),
\end{equation}
where $B_0$ is a normalization constant. It can be shown that the total magnetic charge is
\begin{align}
    Q_m=\frac{1}{4\pi}\int_{\partial \Sigma}F=\frac{1}{4\pi}\int_{\partial \Sigma}d\mathscr{A}=\frac{1}{4\pi}\int_{\partial \Sigma}\mathscr{A}_{\nu,\mu}dx^{\mu}\wedge dx^{\nu}=\frac{1}{4\pi}\int_{\partial \Sigma}\mathscr{A}_{\phi,\theta}d\theta\wedge d\phi=B_0,
\end{align}
where we have chosen the surface $\partial\Sigma$ to be the constant $r$, $t$ surface. The 4-vector potential in Kerr-Schild coordinates can then be obtained through a coordinate transformation as
\begin{equation}
    \mathscr{A}_{\mu}^{KS}=B_0\left(\frac{a \cos\theta}{\Sigma},\frac{a \cos\theta}{\Sigma},0,-\cos\theta\,\frac{r^2+a^2}{\Sigma}\right).
\end{equation}
In the 3+1 formalism, we define $\Phi=-\mathscr{A}_0$, $A_j=\mathscr{A}_j$, $A^{j}=\gamma^{jk}A_k$, then we have $\mathbf{E}=-\nabla\Phi-\partial_t \mathbf{A}$, $\mathbf{B}=\nabla\times\mathbf{A}$. The field components are calculated as follows in the Kerr-Schild coordinates
\begin{align}
    B^r&=\frac{1}{\sqrt{\gamma}}\frac{\partial A_{\phi}}{\partial \theta}=B_0\frac{1}{\sqrt{\gamma}}\frac{(a^2+r^2)(r^2-a^2\cos^2\theta)\sin\theta}{\Sigma^2},\\
    B^\theta&=-\frac{1}{\sqrt{\gamma}}\frac{\partial A_{\phi}}{\partial r}=-B_0\frac{1}{\sqrt{\gamma}}\frac{2a^2r\sin^2\theta\cos\theta}{\Sigma^2},\\
    B^{\phi}&=-\frac{1}{\sqrt{\gamma}}\frac{\partial A_r}{\partial \theta}=B_0\frac{1}{\sqrt{\gamma}}\frac{a(r^2-a^2\cos^2\theta)\sin\theta}{\Sigma^2},\\
    E_r&=-\frac{\partial \Phi}{\partial r}=-B_0\frac{2ar\cos\theta}{\Sigma^2},\\
    E_{\theta}&=-\frac{\partial \Phi}{\partial \theta}=-B_0\frac{a(r^2-a^2\cos^2\theta)\sin\theta}{\Sigma^2},\\
    E_{\phi}&=0.
\end{align}
This is the initial field configuration we use in our simulations.

\section{Equations for the fields and particles}\label{sec:equations}

Suppose the electromagnetic field tensor is $F^{\mu\nu}$ and its dual is ${}^*F_{\mu\nu}=(1/2)\epsilon_{\mu\nu\alpha\beta}F^{\alpha\beta}$. Following \cite{2004MNRAS.350..427K}, we introduce the spatial vectors $\mathbf{B}$, $\mathbf{D}$, $\mathbf{E}$, and $\mathbf{H}$ as
\begin{align}
    B^i&=\alpha {}^*F^{i0}=\frac{1}{2}e^{ijk}F_{jk},\\
    E_i&=\frac{1}{2}\alpha e_{ijk} {}^*F^{jk}=F_{i0},\\
    D^i&=\alpha F^{0i}=\frac{1}{2} e^{ijk} {}^*F_{jk},\\
    H_i&=\frac{1}{2}\alpha e_{ijk}F^{jk}={}^*F_{0i},
\end{align}
where $e^{ijk}=(1/\sqrt{\gamma})[ijk]$, $e_{ijk}=\sqrt{\gamma}[ijk]$ is the Levi–Civita pseudo-tensor of the absolute space. Let $\rho=\alpha J^0$, $j^k=\alpha J^k$, where $J^{\mu}$ is the 4-current, then the Maxwell equations can be written as (we use Heaviside-Lorentz units)
\begin{align}
    \nabla\cdot\mathbf{B}&=0,\\
    \nabla\times\mathbf{E}&=-\frac{\partial \mathbf{B}}{\partial t},\\
    \nabla\cdot\mathbf{D}&=\rho,\\
    \nabla\times\mathbf{H}&=\mathbf{j}+\frac{\partial \mathbf{D}}{\partial t}.
\end{align}
We also have the following constitutive relations
\begin{align}
    \mathbf{E}&=\alpha \mathbf{D}+\pmb{\beta}\times\mathbf{B},\\
    \mathbf{H}&=\alpha \mathbf{B}-\pmb{\beta}\times\mathbf{D}.
\end{align}

The equations of motion for particles are
\begin{align}
    \frac{du_i}{dt}&=-\alpha u^0 \frac{\partial \alpha}{\partial x^i}+u_k \frac{\partial \beta^k}{\partial x^i}-\frac{u_j u_k}{2u^0}\frac{\partial \gamma^{jk}}{\partial x^i}+\frac{q}{m}\left(\alpha D_i+e_{ijk}B^k\frac{\gamma^{jl}u_l}{u^0}\right),\\
    \frac{dx^i}{dt}&=-\beta^i+\frac{\gamma^{ij}u_j}{u^0},
\end{align}
where $u^0=(1/\alpha)\sqrt{\gamma^{ij}u_i u_j+\epsilon}$, and $\epsilon=1$ for massive particles, $\epsilon=0$ for photons.

\section{Calculation of energy flux and averaged quantities}\label{sec:averaging}

In an axisymmetric, stationary spacetime, there are two Killing vectors: the timelike one $\chi^{\mu}$, and the axial one $\eta^{\mu}$. Following \cite{1977MNRAS.179..433B}, we can define conserved flux vectors of energy and angular momentum.
Suppose $T^{\mu\nu}$ is the total stress-energy tensor, and $\xi^{\mu}$ is a Killing vector. From the conservation of energy and momentum,
\begin{equation}
    \left(T^{\mu\nu}\right)_{;\nu}=0,
\end{equation}
and the property of the Killing vector
\begin{equation}
    \xi_{\mu;\nu}+\xi_{\nu;\mu}=0,
\end{equation}
we can obtain
\begin{equation}
    \left(\xi_{\mu}T^{\mu\nu}\right)_{;\nu}=0.
\end{equation}
Therefore, we can define the conserved energy flux as
\begin{equation}
    \mathcal{E}^{\mu}=-T^{\mu\nu}\chi_{\nu}=-\tensor{T}{^\mu_0},
\end{equation}
and the angular momentum flux
\begin{equation}
    \mathcal{L}^{\mu}=T^{\mu\nu}\eta_{\nu}=\tensor{T}{^{\mu}_{\phi}}.
\end{equation}
Here we already used the fact that $\chi^{\mu}=(1,0,0,0)$ and $\eta^{\mu}=(0,0,0,1)$ in Kerr-Schild coordinates. 

The energy flux $\mathcal{E}^{\mu}$ satisfies the following conservation law:
\begin{equation}\label{eq:energy_conservation_diff}
    (\mathcal{E}^{\mu})_{;\mu}=0.
\end{equation}
We can define the energy density 3-form 
\begin{equation}
    {}^*{\pmb{\mathcal{E}}}=\mathcal{E}^{\mu}d^3\Sigma_{\mu}=\pmb{\mathcal{E}}\cdot d^3\pmb{\Sigma}=\mathcal{E}^{\mu}\epsilon_{\mu\alpha\beta\gamma}\pmb{d}x^{\alpha}\wedge\pmb{d}x^{\beta}\wedge\pmb{d}x^{\gamma}/3!,
\end{equation}
where $d^3\Sigma_{\mu}$ is the basis 3-form, and $\epsilon_{\mu\alpha\beta\gamma}$ is the Levi-Civita tensor density in the 4-dimensional spacetime. The energy conservation law (\ref{eq:energy_conservation_diff}) can also be expressed as $\pmb{d}{}^*\pmb{\mathcal{E}}=0$. According to Stokes' Theorem, we can get the corresponding integral conservation law \citep{1973grav.book.....M}
\begin{equation}\label{eq:energy_conservation_integral}
    0=\int_{\mathcal{V}}\pmb{d}{}^*\pmb{\mathcal{E}}=\int_{\partial \mathcal{V}}{}^*\pmb{\mathcal{E}}.
\end{equation}
For example, suppose the spacetime region of interest is bounded by the timelike surfaces $\mathcal{S}_{E}$ at the event horizon and $\mathcal{S}_R$ at radius $R$ (both have positive direction pointing outward), as well as the future $\mathcal{S}_2$ and past $\mathcal{S}_1$ boundaries (both have positive direction pointing toward future), then $\partial\mathcal{V}=\mathcal{S}_2-\mathcal{S}_1+\mathcal{S}_R-\mathcal{S}_E$. The integral conservation law (\ref{eq:energy_conservation_integral}) then states that
\begin{equation}\label{eq:energy_conservation_example}
    E(\mathcal{S}_2)-E(\mathcal{S}_1)=F(\mathcal{S}_E)-F(\mathcal{S}_R),
\end{equation}
where $E(\mathcal{S}_1)=\int_{\mathcal{S}_1}{}^*\pmb{\mathcal{E}}=\int_{\mathcal{S}_1} \mathcal{E}^0\, \sqrt{-g}\,dr\,d\theta\,d\phi$ is the energy in the region at the past boundary $\mathcal{S}_1$ (similarly $E(\mathcal{S}_2)$ is the energy at the future boundary $\mathcal{S}_2$), and $F(\mathcal{S}_R)=\int_{\mathcal{S}_R}{}^*\pmb{\mathcal{E}}=\int_{\mathcal{S}_R}\mathcal{E}^r\sqrt{-g}\,dt\,d\theta\,d\phi$ is the energy flux going out through the outer surface $\mathcal{S}_{R}$ (similarly $F(\mathcal{S}_E)$ is the energy flux going out through the event horizon $\mathcal{S}_{E}$). In Figure \ref{fig:tau5_Lr_Lt}, we calculated the power going through the spherical surface $\mathcal{S}_R$ as
\begin{equation}
    L\equiv \frac{dF(\mathcal{S}_R)}{dt}=\int_{\mathcal{S}_R}\mathcal{E}^r\sqrt{-g}\,d\theta\,d\phi
\end{equation}

The electromagnetic part of the energy flux turns out to be
\begin{align}
    \mathcal{E}_{\rm EM}^0&=\frac{1}{2\alpha}(\mathbf{D}\cdot\mathbf{E}+\mathbf{B}\cdot\mathbf{H}),\\
    \mathcal{E}_{\rm EM}^i&=\frac{1}{\alpha}e^{ijk}E_j H_k\equiv \frac{1}{\alpha} P^i,
\end{align}
where $\mathbf{P}=\mathbf{E}\times\mathbf{H}$. In fact, the Poynting theorem in the $3+1$ formalism can be written as
\begin{equation}
    \frac{\partial}{\partial t}\left[\frac{1}{2}\left(\mathbf{D}\cdot\mathbf{E}+\mathbf{B}\cdot\mathbf{H}\right)\right]+\nabla\cdot(\mathbf{E}\times\mathbf{H})=-\mathbf{E}\cdot\mathbf{j},
\end{equation}
so $U_{\rm EM}\equiv \alpha \mathcal{E}_{\rm EM}^0=(\mathbf{D}\cdot\mathbf{E}+\mathbf{B}\cdot\mathbf{H})/2$ can be regarded as the electromagnetic energy density in the absolute space, and $\mathbf{P}$ is the Poynting flux in the absolute space. As an example, the electromagnetic part of the energy flux $F(\mathcal{S}_R)$ in Equation (\ref{eq:energy_conservation_example}) can be calculated equivalently as follows
\begin{equation}
    F_{\rm EM}(\mathcal{S}_R)=\int_{\mathcal{S}_R}\mathcal{E}_{\rm EM}^r\sqrt{-g}\,dt\,d\theta\,d\phi=\int_{\mathcal{S}_R}P^r\sqrt{\gamma}\,dt\,d\theta\,d\phi.
\end{equation}

For particles, we obtain their contribution to the stress-energy tensor through the distribution function $f(\pmb{x},\pmb{p},t)=dN/(d\mathcal{V}_x d\mathcal{V}_p)$. The first moment of the distribution function in momentum space gives the particle flux
\begin{equation}
    S^{\mu}= \int f p^{\mu} \frac{d\mathcal{V}_p}{p^0}.
\end{equation}
Note that $d\mathcal{V}_p/p^0$ is a Lorentz invariant volume element. The second moment of the distribution function gives the matter stress-energy tensor
\begin{equation}
    T_{\rm MA}^{\mu\nu}=\int f p^{\mu}p^{\nu}\frac{d\mathcal{V}_p}{p^0}.
\end{equation}
In the main text, particle averaged quantity $\langle\Phi\rangle$ is defined as
\begin{equation}
    \langle\Phi\rangle\equiv\frac{\int f \Phi\, d\mathcal{V}_p}{\int f \, d\mathcal{V}_p}=\frac{\int f \Phi\, d\mathcal{V}_p}{S^0}.
\end{equation}
For example, the average Lorentz factor in the FIDO frame $\langle\Gamma\rangle$ is
\begin{equation}
    \langle\Gamma\rangle=\alpha \langle u^0\rangle=\alpha \frac{\int f u^0p^0\, \frac{d\mathcal{V}_p}{p^0}}{S^0}=\alpha \frac{T^{00}}{m S^0}.
\end{equation}
The average energy at infinity $\langle e_{\infty}\rangle$ is
\begin{equation}
    \langle e_{\infty}\rangle=-\langle u_0\rangle=-\frac{\int f u_0p^0\, \frac{d\mathcal{V}_p}{p^0}}{S^0}=-\frac{\tensor{T}{^0_0}}{m S^0}.
\end{equation}
The average radial velocity of the particles is
\begin{equation}
    \langle v^r\rangle=\frac{\int f p^r\frac{d\mathcal{V}_p}{p^0}}{S^0}=\frac{S^r}{S^0}.
\end{equation}

\section{Force-free solution of the monopole magnetosphere}\label{sec:force-free}

\begin{figure*}
    \centering
    \begin{minipage}{0.56\textwidth}
        \includegraphics[width=\textwidth]{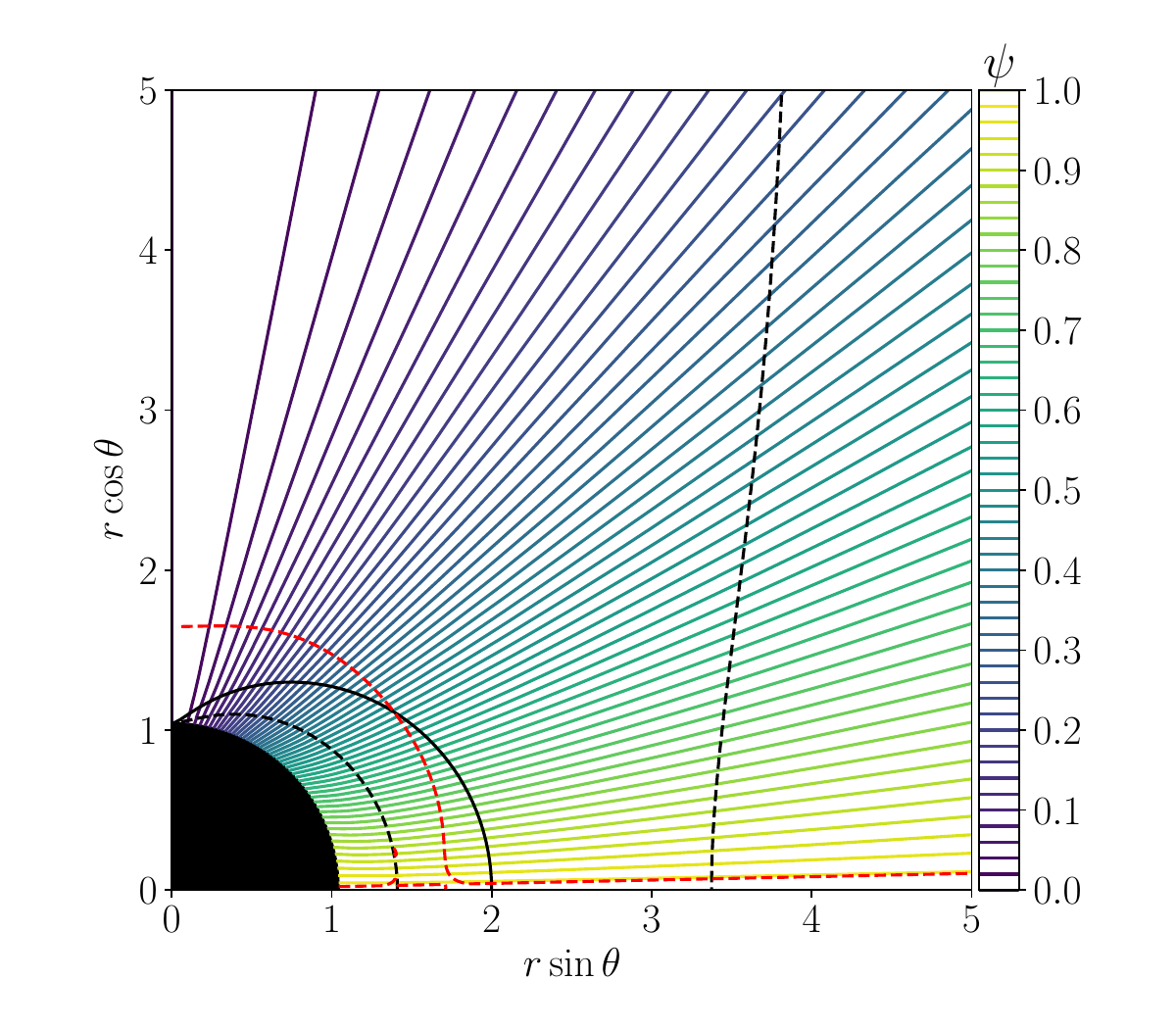}
    \end{minipage}
    \begin{minipage}{0.42\textwidth}
    \vspace{0.2cm}
    \includegraphics[width=0.92\textwidth]{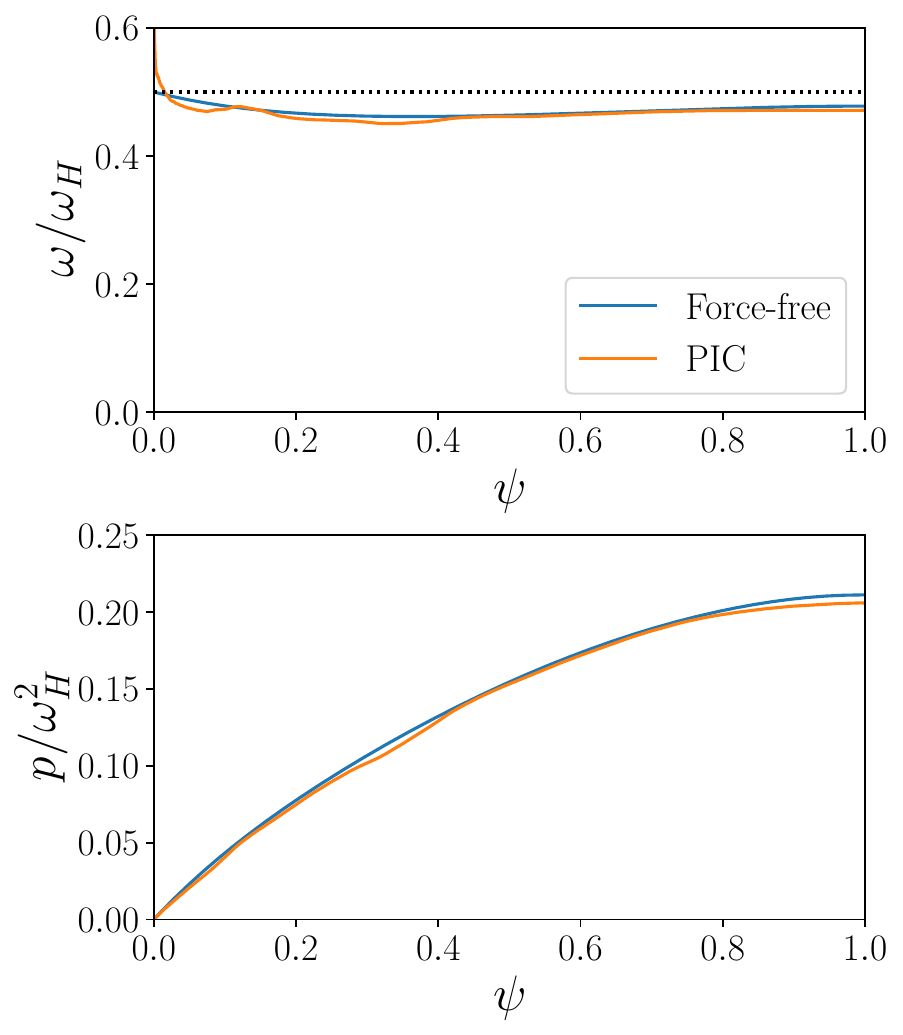}
    \end{minipage}
    \caption{Left: force-free solution of the monopolar magnetosphere for the case $a=0.999$, shown in Boyer-Lindquist coordinates. Colored solid lines are equally spaced contours of the flux function $\psi$. The black solid line is the ergosphere, the black dashed lines are the light surfaces, and the red dashed line is the null surface. Right: The field line angular velocity $\omega(\psi)$ and the outgoing electromagnetic power per unit flux $p(\psi)=\omega(\psi)I(\psi)$ for the force-free solution (blue lines), overlaid with the corresponding values from our fiducial GRPIC simulation during steady state (orange lines). The values from the PIC simulation are taken along a spherical surface at $r=3r_g$. In the top panel, the black dotted line corresponds to $\omega/\omega_H=0.5$.}
    \label{fig:forcefree}
\end{figure*}

We obtain the steady-state, force-free solution of the monopolar magnetosphere by solving the general relativistic Grad-Shafranov equation for the flux function $\psi\equiv A_{\phi}$ (which is the $\phi$ component of the 4-vector potential $A_{\mu}$). In Boyer-Lindquist coordinates, the Grad-Shafranov equation can be written as \citep[e.g.,][]{1977MNRAS.179..433B}
\begin{align}\label{eq:GS}
\frac{B_T(\psi) B_T'(\psi)}{\Delta\sin ^2\theta}&=\frac{1}{\sqrt{-g}}\left[\left(\frac{\sqrt{-g}g^{rr} K}{\Delta \sin ^2 \theta}\psi _{,r}\right)_{,r}+\left(\frac{\sqrt{-g}g^{\theta \theta } K}{\Delta \sin ^2 \theta}\psi _{,\theta }\right)_{,\theta }\right]-\frac{(\nabla \psi )^2 \omega'(\psi) \left(g_{0 \phi }+\omega(\psi)  g_{\phi \phi }\right)}{\Delta  \sin ^2 \theta },
\end{align}
where $K=g_{00}+2  g_{0\phi }\omega(\psi) +g_{\phi \phi }\omega ^2(\psi) $ and $(\nabla \psi )^2=g^{\theta \theta }\left(\partial \psi /\partial \theta \right)^2+g^{rr}\left(\partial \psi /\partial r\right)^2$. The toroidal magnetic field $B_T=\sqrt{-g} g^{r r} g^{\theta \theta } (A_{\theta ,r}-A_{r,\theta }) $ (or equivalently the poloidal current $I=B_T$) and field line angular velocity $\omega$ are functions of $\psi$. A solution to this equation should determine $\psi(r,\theta)$ and the functional forms $B_T(\psi)$, $\omega(\psi)$ at the same time.

We use a relaxation method developed in \citet{2019MNRAS.484.4920Y}, which is similar to earlier works by \citet{2013ApJ...765..113C,2014ApJ...788..186N}. Figure \ref{fig:forcefree} shows the force-free solution of the magnetosphere, as well as the functional form of the field line angular velocity $\omega(\psi)$ and the outgoing electromagnetic power per unit magnetic flux $p(\psi)=\omega(\psi)I(\psi)$. The results are consistent with \cite{2010ApJ...711...50T} and \cite{2013ApJ...765..113C}. We compare the force-free solution with our GRPIC fiducial run during steady state in the right panel of Figure \ref{fig:forcefree}. We see excellent agreement, suggesting that the fiducial run indeed reaches a state very close to force-free.

Note that in Equation (\ref{eq:GS}), if we scale $\psi$ by a constant, $\psi\to\tilde{B}_0\psi$, then $\omega$ should remain the same, while $I\to \tilde{B}_0 I$. Therefore, the total luminosity, $L(\psi)\equiv 2\pi \int p(\psi)\, d\psi$, scales as $L\to \tilde{B}_0^2L$. This confirms that the total outgoing power in the force free solution scales as the square of the magnetic flux threading the horizon.

In the force-free solution, the location of the light surfaces is given by the condition \citep[e.g.,][]{2004MNRAS.350..427K}
\begin{equation}
    g_{00}+2g_{0\phi}\omega+g_{\phi\phi}\omega^2=0,
\end{equation}
where $\omega$ is the angular velocity of the field line. The location turns out to be the same in both Boyer-Lindquist coordinates and Kerr-Schild coordinates. When writing out the fields in the $3+1$ formalism, we find \citep[see also, e.g.,][]{2020PhRvL.124n5101C}
\begin{equation}\label{eq:field_line_angular_velocity}
    \omega=-\frac{E_{\theta}}{\sqrt{\gamma}B^r}.
\end{equation}
In our PIC simulations, we also calculate the angular velocity of field lines in the same manner as in force-free using Equation (\ref{eq:field_line_angular_velocity}).

\bibliography{reference}
\bibliographystyle{aasjournal}

\end{document}